\documentclass[%
preprint,  
 amsmath,amssymb,
 aps,
]{revtex4-1}

\usepackage{graphicx}% Include figure files
\usepackage{dcolumn}% Align table columns on decimal point
\usepackage{bm}% bold math
\usepackage{float}% insert figure at the place where the figure command is put.
\newcommand{\bq}{{\mbox{\boldmath $q$}}}

\newcommand{\bphi}{{\mbox{\boldmath $\phi$}}}

\newcommand{\bra}{\langle}
\newcommand{\ket}{\rangle}

\newcommand{\bd}{\begin{displaymath}}
\newcommand{\ed}{\end{displaymath}}

\newcommand{\hatq}{\hat{q}}
\newcommand{\hatQ}{\hat{Q}}

\newcommand{\bea}{\begin{eqnarray}}
\newcommand{\eea}{\end{eqnarray}}
\newcommand{\beas}{\begin{eqnarray*}}
\newcommand{\eeas}{\end{eqnarray*}}
\newcommand{\be}{\begin{equation}}
\newcommand{\ee}{\end{equation}}
\def\rmcc{{\rm cc}}
\def\rmcs{{\rm cs}}
\def\rmsc{{\rm sc}}
\def\rmss{{\rm ss}}

\def\rmc{{\rm c}}

\def\vec0{\mbox{\boldmath $0$}}

\begin{document}
%\preprint{APS/123-QED}

\title{
  Correspondence between Phase Oscillator Network and
  Classical XY Model with the same random and frustrated interactions}
%  full paper jpsj3/REVTex10$\_$JPSJ/v10.uezu.full.tex
%}
\author{Tomoyuki Kimoto$^1$} \email{kimoto@oita-ct.ac.jp}
\author{Tatsuya Uezu$^2$}\email{uezu@ki-rin.phys.nara-wu.ac.jp} 

\affiliation{$^1$National Institute of Technology, Oita College, Oita 870-0152, Japan}%
 \affiliation{
$^2$Graduate School of Humanities  and Sciences,
Nara Women's University, Nara 630-8506, Japan}

\date{\today}% It is always \today, today,
%osci.plus.mexican/pre/uezu/v11.touka.genri.tex

\begin{abstract}

%\abst{%
%12345678911234567892123456789312345678941234567895  
We study correspondence between a phase oscillator
network with distributed natural frequencies and
a classical XY model at finite temperatures with
the same random and frustrated interactions used in the
Sherrington-Kirkpatrick model. We perform numerical
calculations of the spin glass order parameter $q$ and the
distributions of the local fields.  As a result, we find that
the parameter dependences of these quantities in both models agree
 fairly well 
if parameters are normalized by using the previously
obtained  correspondence relation between two models
with the same  other types of interactions.
Furthermore, we numerically calculate  several quantities such as
%the time evolution of the instantaneous spin glass order parameter $q$ and its probability density 
the time evolution of the instantaneous local field 
in the phase oscillator network in order to study the roles of synchronous and asynchronous
oscillators.  We also study the self-consistent equation of the local fields in the oscillator
 network and XY model derived by the mean field approximation.

\begin{description}
\item[PACS numbers]
%\verb+\pacs{#1}+ command.
%\pacs{
%05.45.Xt,05.20.-y,05.45.-a,87.10.Rt
%}% PACS, the Physics and Astronomy
05.45.Xt \ % Synchronization; coupled oscillators,
%\keywords{Suggested keywords}%Use showkeys class option if keyword %display desired
05.45.-a \ % Nonlinear dynamics and chaos,                          
05.20.-y \ % Classical statistical mechanics,
%%%%87.10.Rt \ % Monte Carlo simulations
%May be entered using the 
%use the optional argument of the \verb+\item+ command to give the category of ach item. 
%osci.plus.mexican/revtex4-1/tex/latex/revtex/rev5.prl.tex
\end{description}
\end{abstract}

\maketitle

% \email{kimoto@oita-ct.ac.jp}
%\email{uezu@ki-rin.phys.nara-wu.ac.jp} 
%Equivalence between self-consistent equation for phase oscillator network 
%and saddle-point equation for classical XY model}
%\thanks{osci.plus.mexican/revtex4-1/tex/latex/revtex/prl4.tex}

%osci.plus.mexican/pre/uezu/v11.touka.genri.tex

\section{Introduction}
The classical XY model which describes  magnetism has been studied and
 a lot of phase transition phenomena have been found\cite{domb.green}.
 On the other hand, there are a lot of synchronization phenomena in nature
 such as the circadian rhythms, beat of heart, collective firing of fireflies, and so on\cite{saunders,biological.clocks}.
For such synchronization phenomena, the phase oscillator model
which describes oscillations only by phases has been proposed\cite{winfree}, and
 the synchronization-desynchronization phase transition point
has been analytically obtained
in the case of the uniform infinite-range interaction\cite{kuramoto-1}.
 The models which are described only by phases are not special in the sense that
 the differential equations for phases are derived when
   nonlinear differential equations which exhibit limit cycle oscillations
 are weakly coupled \cite{kuramoto-book}. 
  The phase oscillator model with the uniform infinite-range interaction is called
 the Kuramoto model. Since Kuramoto proposed the model,
there have been many extensions of the model, and many interesting phenomena such
  as chimera states and the synchronization due to common noises have been found, and
  attempts to identify a dynamical system from experimental data
   have been made\cite{rev.mod.phys}.\par
In the XY model and the phase oscillator network with the same interaction,
the order parameters are the same,
 and it is trivial that the XY model with zero temperature and
the phase oscillator network with uniform natural frequencies are equivalent, 
 but previously no relations between these two models have been found beyond this. 
 A few years ago, for a class of infinite-range interactions,
we found  the correspondence between
the XY model with non-zero temperature and the phase oscillator network
 with distributed natural frequencies\cite{uezu.kimoto.etal}.
 Specifically, temperature $T$ in the XY model corresponds to the width of
distribution of natural frequencies in the oscillator network, {\it e.g.,} 
$T$ corresponds to $\sqrt{2/\pi}\sigma $
where $\sigma$ is the standard deviation when
the distribution is Gaussian.
The integration kernels for the saddle point equations (SPEs) for
 the XY model and the self-consistent equations (SCEs) for the phase oscillator network
   correspond as well.
 Furthermore,  for several interactions, there exists one-to-one correspondence
 between solutions for  both models, and thus, it is found that the critical exponents
 are the same in both models\cite{in preparation}.\par
In what situations  correspondence between the two models holds is a very interesting
theme.  So far, it has been found that   correspondence holds when
 a few order parameters exist and their SPEs and SCEs are derived 
 for a class of infinite-range interactions with or without randomness and without frustration.
 We have been studying whether  correspondence between the two models exists or not
 for the  interactions for which the SPEs and/or SCEs are  not derived.
 In this paper, in particular,
 we numerically  study  
 random frustrated interactions which were used in the
 Sherrington-Kirkpatrick (SK) model\cite{sk}. We call it the Sherrington-Kirkpatrick (SK)
 interaction in this paper. 
 It is well known that the SK model exhibits
 the spin glass phase for some parameter range.
 In the spin glass phase, the total magnetization is zero, but locally
 each spin is frozen and has non-zero local magnetization.
 The spins with continuous $n$ components are also studied in Ref. 10),
 and the SPEs are derived and the spin glass phase is obtained.
On the other hand, for the phase oscillator network,
more than two decades ago, a numerical study for the SK interaction was performed by Daido
 and non-trivial behaviors were obtained \cite{Daido-1992}.
 That is, the quasientrainment (QE) state was observed, in which the substantial frequency
 for each oscillator is very small, but phases between two such oscillators diffuse
 slowly. Furthermore, the distribution of the local fields (LFs) undergoes a phase transition
that the peak position of the distribution changes from zero to non-zero value as
a parameter changes and this is called the volcano transition. \par
 In this paper, we perform numerical calculations and study
 the spin glass order parameter $q$ and distributions of LFs in both models.
  In addition, in order to study the roles of synchronous and asynchronous oscillators
 in the phase oscillator network, we numerically calculate  several quantities such as
 the time evolution of
 %the instantaneous spin glass order parameter $q$ and its probability
 the phases of oscillators and local fields, and derive
 the SCEs of the LFs assuming that only the synchronous oscillators exist.
 Similarly,  in the XY model, by using the naive mean-field approximation, we derive
 the SCEs of the LFs.  We compare theoretical results with  numerical ones in both models.\par
  The structure of this paper is as follows. In sect. 2, we formulate the problem and
 describe the SPEs. 
 In sect. 3, we show the results of numerical simulations.
Summary and discussion are given in sect. 4.
 In Appendix A, we derive the disorder averaged  free energy per spin and the SPEs
 under the ansatz of the replica symmetry in the XY model.
 
\section{Formulation}
% Now, let us  formulate two systems.
The classical XY model consists of $N$ XY spins
$X_j = (\cos \phi_j, \sin \phi_j), (j=1,\cdots,N$),
where $\phi_j$ is the phase of the $j$th XY spin.
The Hamiltonian $H$  is given by 
\begin{equation}
  H = -\sum_{j < k}^{N} J_{jk} \cos(\phi_j-\phi_k) \label{eq:H},
\end{equation}
where $J_{jk}$ is the interaction between the $j$th and $k$th XY spins. 
 On the other hand, in the  phase oscillator network, 
 each oscillator is described by a phase.
  Let $\phi_j$ be the phase of the $j$th phase oscillator.
 The evolution equation for $\phi_j$ is 
 given by 
\begin{equation}
  \frac{d \phi_j}{d t} = \omega_j + \sum_{k=1}^{N} J_{jk} \sin(\phi_k-\phi_j), \label{eq:TimeUpdate}
\end{equation}
where $J_{jk}$ is the interaction from the $k$th to $j$th phase oscillators,
the constant $\omega_j$ is natural frequency. We assume that  
$\omega_j$ is a  random variable generated from the
probability density function $g(\omega)$. We assume
that $g(\omega)$ is one-humped and  symmetric with respect to
 its center $\omega_0$. In this paper, as $g(\omega)$ we adopt  the Gaussian distribution
with mean 0 and standard deviation $\sigma$, ${\mathcal{N}}(0,\sigma^2)$.
We assume both systems have the following SK interaction in common:
\begin{equation}
  J_{jk}  = \frac{J}{\sqrt{N}} z_{jk}, %\;\;\;  z_{jk} \sim {\mathcal{N}}(0,1),
  \label{eq:JJK} \\
\end{equation}
where $z_{jk}$ is a random variable obeying the Gaussian distribution ${\mathcal{N}}(0,1)$.
Moreover, we assume  $J_{jj}=0$ and $J_{jk}=J_{kj} ( j \ne k)$.

Now, by using the replica method, we derive the saddle point equations (SPEs) for
the XY model, which is originally obtained in Ref.\cite{sk}.

Firstly, in the XY model, we define the following spin glass order parameter $q$: 
\begin{equation}
  q = \mbox{Max} \left( \left| \frac{1}{N} \sum_{j=1}^N
  e^{i \left( \phi_j^\alpha-\phi_j^\beta \right) } \right|,
  \left| \frac{1}{N} \sum_{j=1}^N
  e^{i \left( \phi_j^\alpha-(-\phi_j^\beta) \right) } \right| \right),  \label{eq:q}
\end{equation}
where $i=\sqrt{-1}$,
$\phi_j^\alpha \ (1 \le j \le N)$ and $\phi_j^\beta \ (1 \le j \le N)$
are phases of two replicas $\alpha$ and $\beta$ that have the same interaction $\{J_{jk}\}$.
%$\mbox{Max}(\cdot)$ in Eq. (\ref{eq:q}) is the function that returns
%the larger value from two arguments.
The first argument is calculated by the phase difference between 
$\phi_j^\alpha$ and $\phi_j^\beta$,
and the second argument is calculated by the phase difference between 
$\phi_j^\alpha$ and $-\phi_j^\beta$.
Since the Hamiltonian (\ref{eq:H}) has the reversal symmetry, that is it is invariant
 under the reversal of signs of phases $\{\phi_j\} \to \{-\phi_j\}$, 
  we calculate the summation for the reversal phase $-\phi_j^\beta$ shown in the second argument.
  Since we set $J_0=0$ and $J=1$, then  $q>0$ when the system  is in  the spin glass state,
and $q=0$ when it is in the paramagnetic state.

Introducing $n$ replicas, we define the following order parameters.
For $\alpha< \beta$, 
\bea
&& q_{\rmcc}^{\alpha \beta}=\frac{1}{N}\sum_i \cos \phi_i ^\alpha \cos \phi_i ^\beta, \
q_{\rmss}^{\alpha \beta}=\frac{1}{N}\sum_i \sin \phi_i ^\alpha \sin \phi_i ^\beta, 
\eea
and for $\alpha \ne \beta$,
\bea
&&q_{\rmcs}^{\alpha \beta}=\frac{1}{N}\sum_i \cos \phi_i ^\alpha \sin \phi_i ^\beta,
\eea
and for $\alpha=1, \cdots,n$, 
\bea
&& Q_{\rmcc}^{\alpha}=\frac{1}{N}\sum_i \cos^2 \phi_i ^\alpha, \ 
Q_{\rmss}^{\alpha}=\frac{1}{N}\sum_i \sin ^2 \phi_i ^\alpha, \
Q_{\rmcs}^{\alpha}=\frac{1}{N}\sum_i \cos \phi_i ^\alpha \sin \phi_i ^\alpha.\nonumber 
\eea
By using the standard recipe, we obtain the disorder averaged  free energy per spin
$\bar{f}=-\lim _{N \to \infty} (\beta N)^{-1} \overline{\log Z}$
by the replica method. Here, $\overline{ \cdots  }$ implies the average over $\{J_{ij} \}$.
Assuming the replica symmetry, we obtain
\bea
   \bar{f}_{\rm RS} &=& -\frac{1}{\beta}
   \biggl\{
   \frac{\beta ^2 J^2}{4} (
   q_{\rmcc}^2+q_{\rmss}^2+2q_{\rmcs}^2
 -Q_{\rmcc}^2-Q_{\rmss}^2 -2Q_{\rmcs}^2)\nonumber \\
&&  +\int Dx \int Dy \log \int d \phi M(\phi | x,y) \biggr\}, \\
 M(\phi| x,y)&=& \exp\biggl[
 \frac{\beta ^2 J^2}{2} (  Q_{\rmcc}  - q_{\rmcc}) \cos ^2 \phi+
   \frac{\beta ^2 J^2}{2} (  Q_{\rmss}  - q_{\rmss}) \sin ^2 \phi \nonumber \\
&&+ \beta ^2 J^2 (  Q_{\rmcs}  - q_{\rmcs}) \sin \phi  \cos \phi \nonumber \\
&& + \beta J
\sqrt{\frac{q_{\rmcc} q_{\rmss}-(q_{\rmcs})^2}{q_{\rmss}}}     
       \cos \phi \ x
       + \beta J \biggl(\frac{q_{\rmcs}}{\sqrt{q_{\rmss}}}\cos \phi
       +\sqrt{q_{\rmss}} \sin \phi \biggr)y \biggr].
 \eea
 From this, we obtain the following SPEs.
 \bea
 && Q_{\rmcc}=[\bra \cos ^2 \phi \ket], Q_{\rmss}=[\bra \sin ^2 \phi \ket]=1-Q_{\rmcc}, \
 Q_{\rmcs}=[\bra \sin \phi  \cos \phi \ket],\\
 && q_{\rmcc}=[\bra \cos \phi \ket^2], q_{\rmss}=[\bra \sin \phi \ket^2], \
 q_{\rmcs}=[\bra \sin \phi \ket \bra \cos \phi \ket],\\
 && [\cdots]\equiv \int Dx \int Dy \cdots, \ \bra \cdots \ket \equiv \frac{\int d \phi M(\phi|x,y)
   \ \cdots }{\int d \phi M(\phi|x,y)}.
 \eea
 See Appendix A for the derivation.
 \if0
 The spin glass order parameter $q$ is defined by
 \bea
 q =\mbox{ Max } \{
 \left| \frac{1}{N} \sum_{j=1}^N  e^{i \left( \phi_j^\alpha-\phi_j^\beta \right) } \right|,
 \left| \frac{1}{N} \sum_{j=1}^N  e^{i \left( \phi_j^\alpha+\phi_j^\beta \right) } \right|
 \}. 
  \label{sgq}
  \eea
  \fi
  $q$ defined by eq. ({\ref{eq:q}) is rewritten by using these quantities as
  \bea
  q &=&\mbox{ Max } \{
  \sqrt{ (q_{\rmcc}^{\alpha \beta}+q_{\rmss}^{\alpha \beta})^2
    +  (q_{\rmsc}^{\alpha \beta}-q_{\rmcs}^{\alpha \beta})^2},
    \sqrt{ (q_{\rmcc}^{\alpha \beta}-q_{\rmss}^{\alpha \beta})^2
    +  (q_{\rmsc}^{\alpha \beta}+q_{\rmcs}^{\alpha \beta})^2} \}.
  \label{q2}
  \eea
  We found $Q_\rmcc =Q_\rmss=\frac{1}{2}, Q_\rmcs=0$ by the Markov Chain
   Monte Carlo simulations (MCMCs).  
  As for $q$s, we found  several relations among them
  depending on samples.
  %$q_\rmcc \simeq q_\rmss$ and $q_\rmsc \simeq - q_\rmsc >0$
%  hold for several parameters.
  Assuming $Q_\rmcc =Q_\rmss=\frac{1}{2}, Q_\rmcs=0$,
  we solved the SPEs for $q_\rmcc, q_\rmss$, and $q_\rmcs$,
  and obtained $q_\rmcc \simeq q_\rmss$ and $q_\rmcs \simeq q_\rmsc  \simeq 0$.
  By assuming  $q_\rmcc = q_\rmss$ and $q_\rmcs =q_\rmsc = 0$,
  we obtain 
  \bea
q&=&2q_{\rmcc}.
\eea
In Appendix A, we prove that
$q$ obeys the same equation as that derived by Sherrington and Kirkpatrick\cite{sk},
\bea
 q&=& 1  -      \frac{k_{\rm B} T}{ J} \sqrt{\frac{2}{q}}
  \int_0 ^\infty dr  r^2  e^{-\frac{1}{2}r^2} 
  \frac{I_1(\frac{J}{k_{\rm B}T} \sqrt{\frac{q}{2}}r)}
{I_0(\frac{J}{k_{\rm B}T} \sqrt{\frac{q}{2}}r)},
\eea
where %$q=q_{\rmcc}+q_{\rmss}=[\bra \cos(\phi^\alpha - \phi^\beta) \ket]$, 
$\beta=\frac{1}{k_{\rm B}T}$,   $k_{\rm B}$ is
the Boltzmann constant, and $I_n$ is the $n$th modified Bessel function.
 The critical temperature is $T_{c}=J/2$ below which
 the spin glass phase appears.

\section{Numerical simulation}
Here, we show numerical results. In this paper, we set  $J_0=0$ and $J=1$ and then 
 $T_c=0.5$.
%\subsection{Comparison of spin glass order parameter $q$}
\subsection{Spin glass order parameter $q$}
\subsubsection{XY model}
Now, let us explain our method of numerical calculations.
 We use the replica exchange Monte Carlo (REMC) method. 
 We prepared 48 sets and 96 sets  of  temperature for $N=100$ and 500, respectively,
  and a replica is assigned  to each  temperature.  We call it a temperature replica.
 The temperature $T$ ranges from 0.02 to 0.96 with the increment
 $\Delta T=0.02$ for $N=100$ and $\Delta T=0.01$ for $N=500$, respectively. 
 The initial values of $\{\phi_j\}$ of all  replicas were set to values in $[0, 2 \pi)$
  randomly.  In order to calculate $q$, we prepare another  set of replicas.
  Two sets of replicas are denoted by $\alpha$ and  $\beta$, respectively.
  For $N=100$ (500), 
  we exchange temperature replicas every 5000 (1000) Monte Carlo sweeps (MC sweeps).
   One MC sweep corresponds to $N$ updates of spins.
  The number of exchanges is 10000. After 500 exchanges, 
  at each temperature, we calculate the time average of $q$
  using 100 sets of phases of XY spins 
  for the last 100 MC sweeps during 5000 and 1000 MC sweeps for $N=100$ and 500,
   respectively. We denote this average by  $\bar{q}$.
  Then we take the average of $\bar{q}$
   over 9500 exchanges, which we regard as the thermal average $\bra q \ket$.
   At each temperature,  the sample average of $\bra q \ket $
   and its  standard deviation  are calculated. % We denote this by $[q]$. 
   The number of samples is 30 and 5 for $N=100$ and  for $N=500$, respectively.
   We show the results of the temperature dependence of  $q$ 
   in Fig. \ref{fig:XY_q-T}(a) for $N=100$ and  
 in Fig. \ref{fig:XY_q-T}(b) for $N=500$.
%   in Fig. \ref{fig1}(a) for $N=100$ and  
% in Fig. \ref{fig1}(b) for $N=500$.
 The solid curves are the theoretical results at the thermodynamic limit of $N=\infty$.
 The theoretical curves look straight, but they are slightly curved.
 The black circles are the sample average of $q$ and the error bars are
the standard deviation. 
 The theoretical curves and the computer simulation
 results almost agree with each other at $T<0.3$ for $N=100$, and at $T<0.4$ for $N=500$,
  respectively.
Therefore, it is expected that the agreement between
the theoretical curves and the simulation results
 becomes better as $N$ is increased, and the critical temperature
 will be $T_c=0.5$ which is the theoretical result.
     %fig1
 \begin{figure}[H]
 \begin{center}
   \begin{tabular}{cc}
     (a) & (b) \\
	\includegraphics[width=4.0cm]{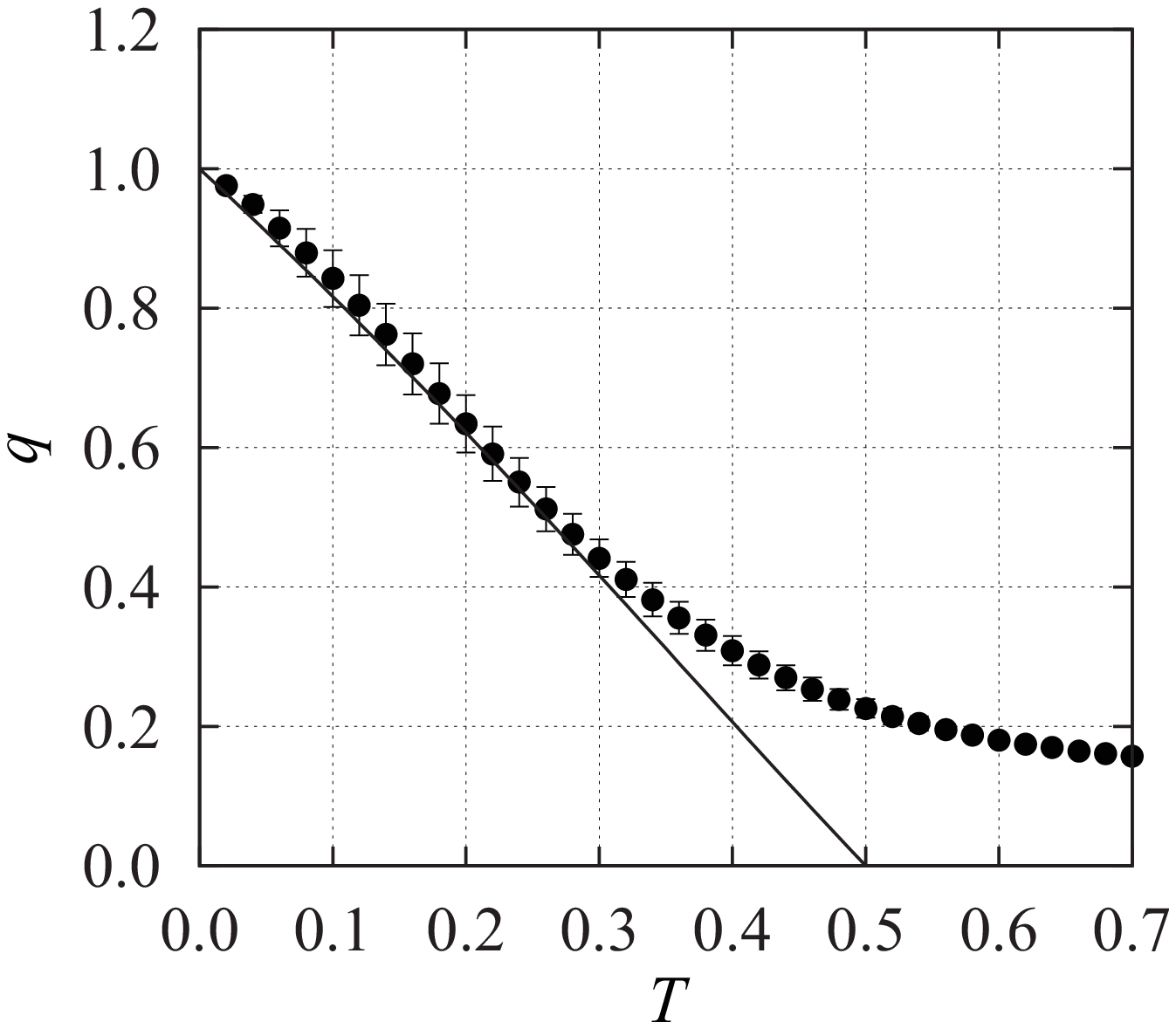}  &
	\includegraphics[width=4.0cm]{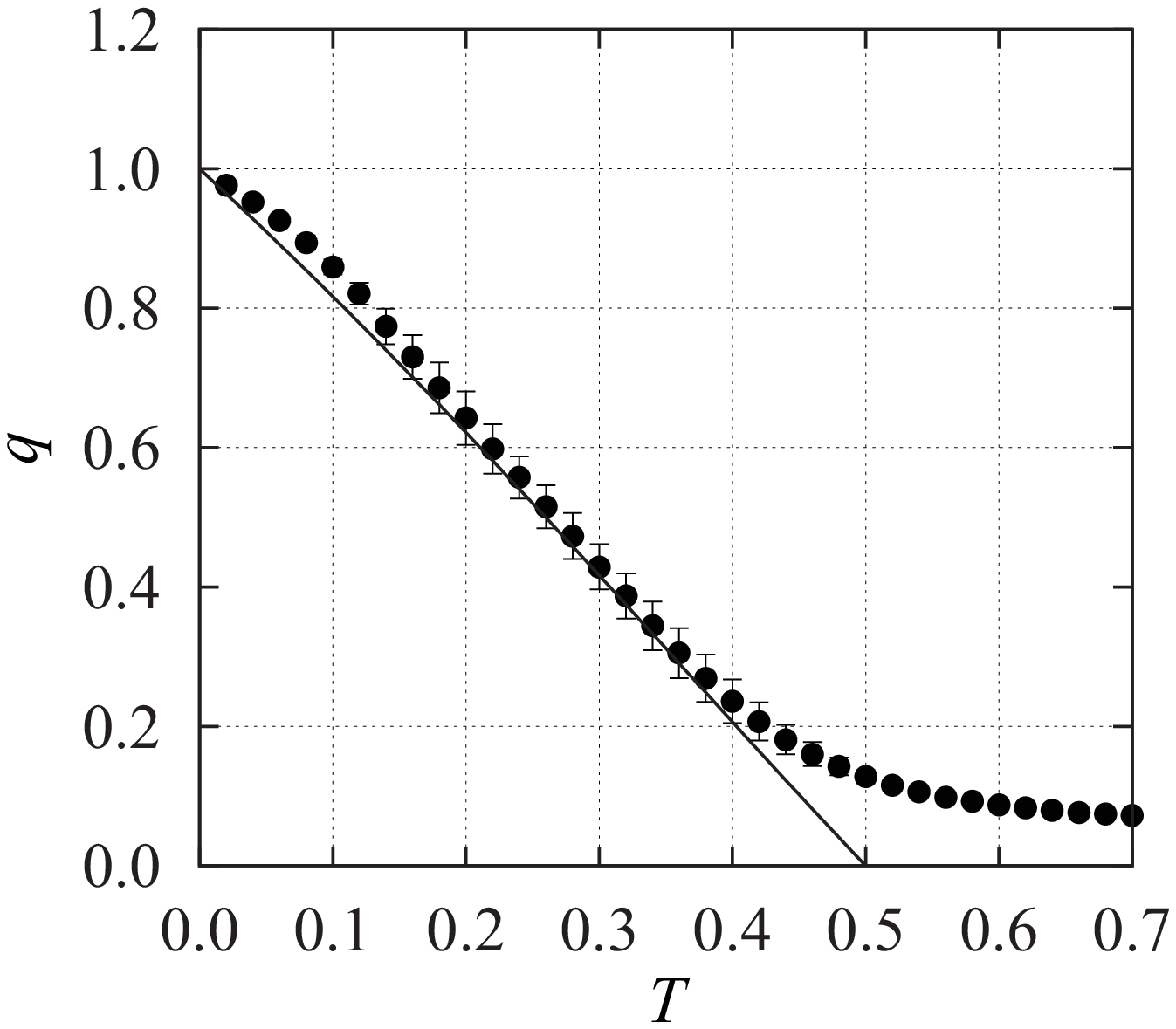} \\
%	\includegraphics[width=4.0cm]{SK-XY_q-t_N=100_NR=48-map(typeB).eps}  &
%	\includegraphics[width=4.0cm]{SK-XY_q-t_N=500_NR=96-map(typeB)-2.eps} \\
%	\small(a) $N=100$  &  \small(b) $N=500$ 
    \end{tabular}
    \caption{Temperature dependence of sample average of $q$ in XY model.
     (a) $N=100$, (b) $N=500$.}
  \label{fig:XY_q-T}
%  \label{fig1}    
 \end{center}
\end{figure}
\subsubsection{Phase oscillator network}
% Next, we calculate $q$ in the  phase oscillator network.
We adopt the same definition of $q$ by Eq. (\ref{eq:q}) as in the XY model.
 In order to guarantee the same
 reversal symmetry as in the XY model, we generate $\omega_j$ for $j=1,2,\cdots,N/2$,
 and set $\omega_j=-\omega_{j-N/2}$ for $j=N/2+1,N/2+2,\cdots,N$.
 The computer simulation was carried out by the following method.
  In order to integrate Eq. (2) numerically, we adopt the Euler method with time increment
  $\Delta t=0.02$.  
Since the Hamiltonian is not defined for the phase oscillator network,
it is impossible to use the REMC method. 
Therefore, in analogy to the simulated annealing method, 
the relaxation calculation was carried out while gradually lowering $\sigma$ from
$ \sqrt{\pi/2}$ to $0$ with the increment
$\Delta \sigma=0.01 \sqrt{\pi/2}$ for $N=100$,
$\Delta \sigma=0.005 \sqrt{\pi/2}$ for $N=200$, and 
 $\Delta \sigma=0.001 \sqrt{\pi/2}$ for $N=500$. 
 In this paper, we also call this the simulated annealing method.
 At each $\sigma$, we  evolve the system until $t=800$ 
 and calculate the time average of $q$
 using phases of oscillators starting from  $t= 501$ to $t= 800$
 with time interval 1.
  We denote this by  $\bar{q}$. 
 At each $\sigma$, 
 the sample average of $\bar{q}$, %which we denote by $[q]$,
 and the standard deviation over samples are calculated.
For this simulated annealing method, $\omega_j \ (1\le j \le N)$ is not generated
for every $\sigma$. Instead, firstly, 
$\omega_j $ with $\sigma=1$ is generated according to
${\mathcal{N}}(0,1)$. We denote it $\omega_{j,0}$. 
 Then, $\omega_j $ with $\sigma (\ne 1)$ is defined as $\sigma \omega_{j,0}$. 
The initial values of $\phi_j \ (1\le j \le N)$ at the beginning  of the 
simulated annealing method are chosen randomly from  $[0, 2\pi)$.
In the simulated annealing method, there may be cases that the relaxed
state is captured at a local minimum.  
In order to  judge whether the relaxed state reached
the global minimum at $\sigma=0$, 
 we used the fact that the phase oscillator network with $\sigma=0 $
and the XY model with $T=0$ are the same model.
Concretely, we used the following method.
 We prepared  the same interaction for both models. 
In the oscillator network, we chose two replicas with
$q \simeq 1$ at $\sigma \sim 0$  obtained by the simulated annealing method.     
Then, 
 we  calculated $q$ 
 using  $\phi_j \ (1\le j\le N)$ of one of two replicas of 
 the phase oscillator network at $\sigma \sim 0 $
  and $\phi_j \ (1\le j\le N)$ of the XY model at $T=0.02$
 obtained by the REMC method. 
 If $q>0.99$, it was judged that the two replicas in the
 oscillator network  reached the global minimum.
 % We generated 30 sets of interactions, and
 By this procedure,  we obtained 100 ($N=100$), 100 ($N=200$),
 and 15 ($N=500)$ pairs of replicas which reached the global minimum
  at $\sigma=0$.
  From the thus obtained  $\bar{q}$s
   for $\sigma > 0$, we calculated the sample average
   of $q$  and the standard deviation. %  $[q]$\\
%  $ 30 (N=100), x(N=200), y(N=500)$ ?\\
%    ----------------------------\\
\if0
    In the simulated annealing method, as the system size increases,
 the simulation time required to reach the global minimum drastically  increases.
 $N=100$ was the limit size that could reach the global minimum in a
 realistic time even if we use a high-speed computer.
\fi
% -------------------------------------\\
% Insert here the description for $N=500$.\\
% -------------------------------------\\
  In  Fig. \ref{fig:OSC_q-sigma}, we display the $\sigma$ dependence of the sample average
  of $q$ with its standard deviation. 
  The solid curve is obtained by the theoretical formula of $q$
   for the XY model by setting  $\sigma=T \sqrt{\pi/2}$.
   For $\sigma<0.4$ when $N=100$, $\sigma<0.2$ when $N=200$, and
   $\sigma<0.17$ when $N=500$,  the  theoretical curve and the  simulation results almost agree.
   However, contrary to our expectation, 
   %It is expected that as the system size increases,
   as the system size increases, 
   the coinciding range of the theoretical curve and the  simulation results decreases.
   The reason for this is considered that $\phi_j$ behaves intermittently in time as we show later.
   In order to observe the averaged behavior, we introduce the following definition of $q_{\rm av}$
   for two replicas $\{\phi_j ^{\alpha}\}$ and $\{\phi_j ^{\beta}\}$.
   \bea
   q_{\rm av} &=& {\rm Max}
   \biggl(
   \frac{|\sum_{j=1}^N   \bar{A}_j^\alpha \bar{A}_j^\beta e^{i(\bar{\phi}_j ^\alpha - \bar{\phi}_j ^\beta)} |}
{\sum_{j=1}^N\bar{A}_j^\alpha \bar{A}_j^\beta}, 
   \frac{|\sum_{j=1}^N   \bar{A}_j^\alpha \bar{A}_j^\beta e^{i(\bar{\phi}_j ^\alpha + \bar{\phi}_j ^\beta)} |}
        {\sum_{j=1}^N\bar{A}_j^\alpha \bar{A}_j^\beta} \biggr), \\
        && \bar{A}_j^\alpha e^{i\bar{\phi}_j ^\alpha}=\frac{1}{T_s}\sum_t ^{T_s}  e^{i \phi_j ^\alpha(t)}, \ \
 \bar{A}_j^\beta e^{i \bar{\phi}_j ^\beta}=\frac{1}{T_s}\sum_t ^{T_s}  e^{i\phi_j ^\beta(t)},
 \eea
 where $T_s=300$.
 The numerical results are shown in Fig. \ref{fig:OSC_qav-sigma}
 for $N=100, 200$, and $N=500$.  From this, we note that
 the  order parameter $q_{\rm av}$ for the time averaged phases
 agree with the theoretical curve fairly well,
 and as $N$ increases 
    the coinciding range of the theoretical curve and the  simulation results increases, and 
    the critical parameter will be  $\sigma_c=T_c \sqrt{\pi/2}$ when $N=\infty$.
    %fig2
\begin{figure}[H]
  \begin{center}
    \begin{tabular}{ccc}
      (a) & (b) & (c) \\
   \includegraphics[width=4.0cm]{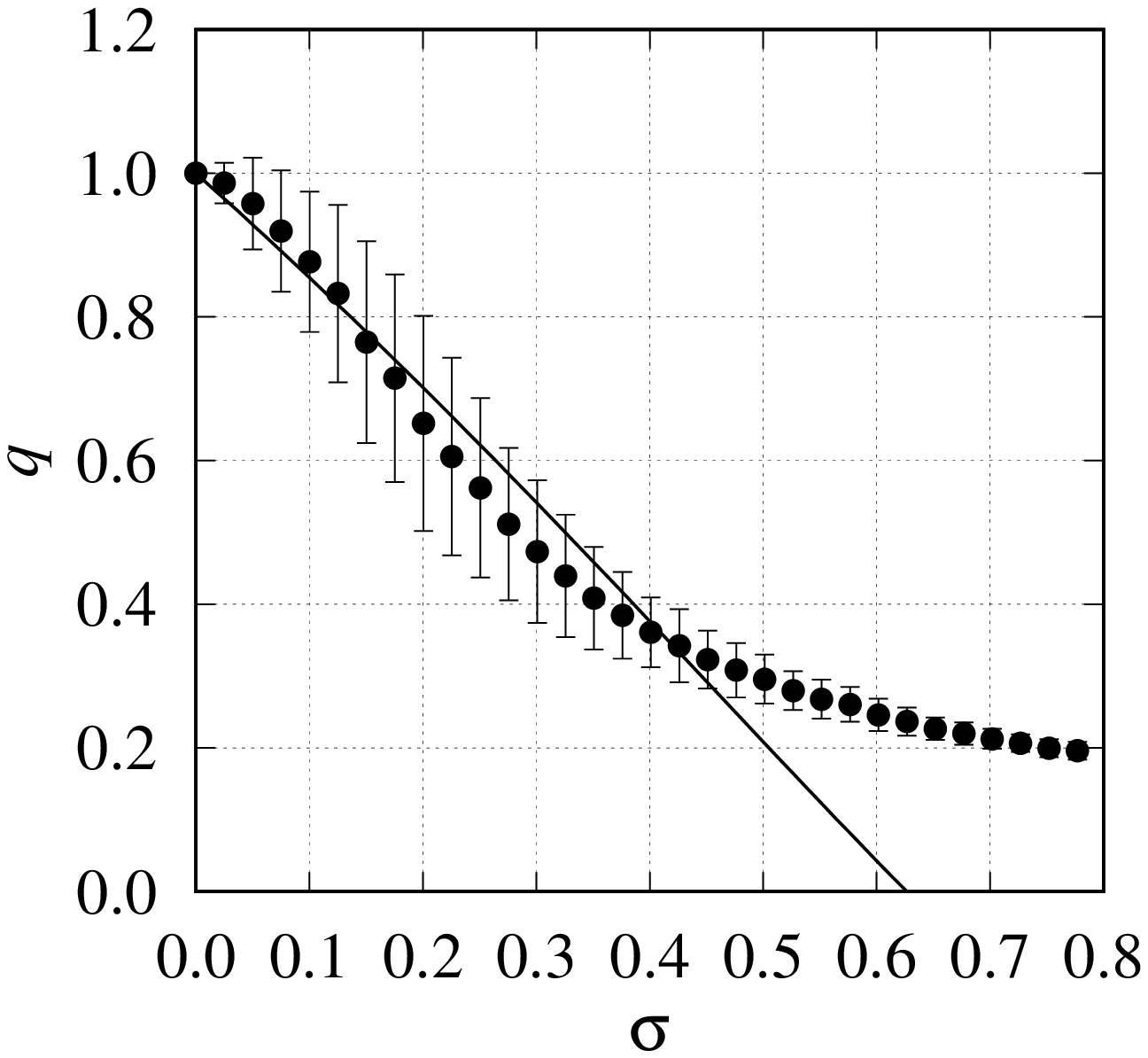} &
   \includegraphics[width=4.0cm]{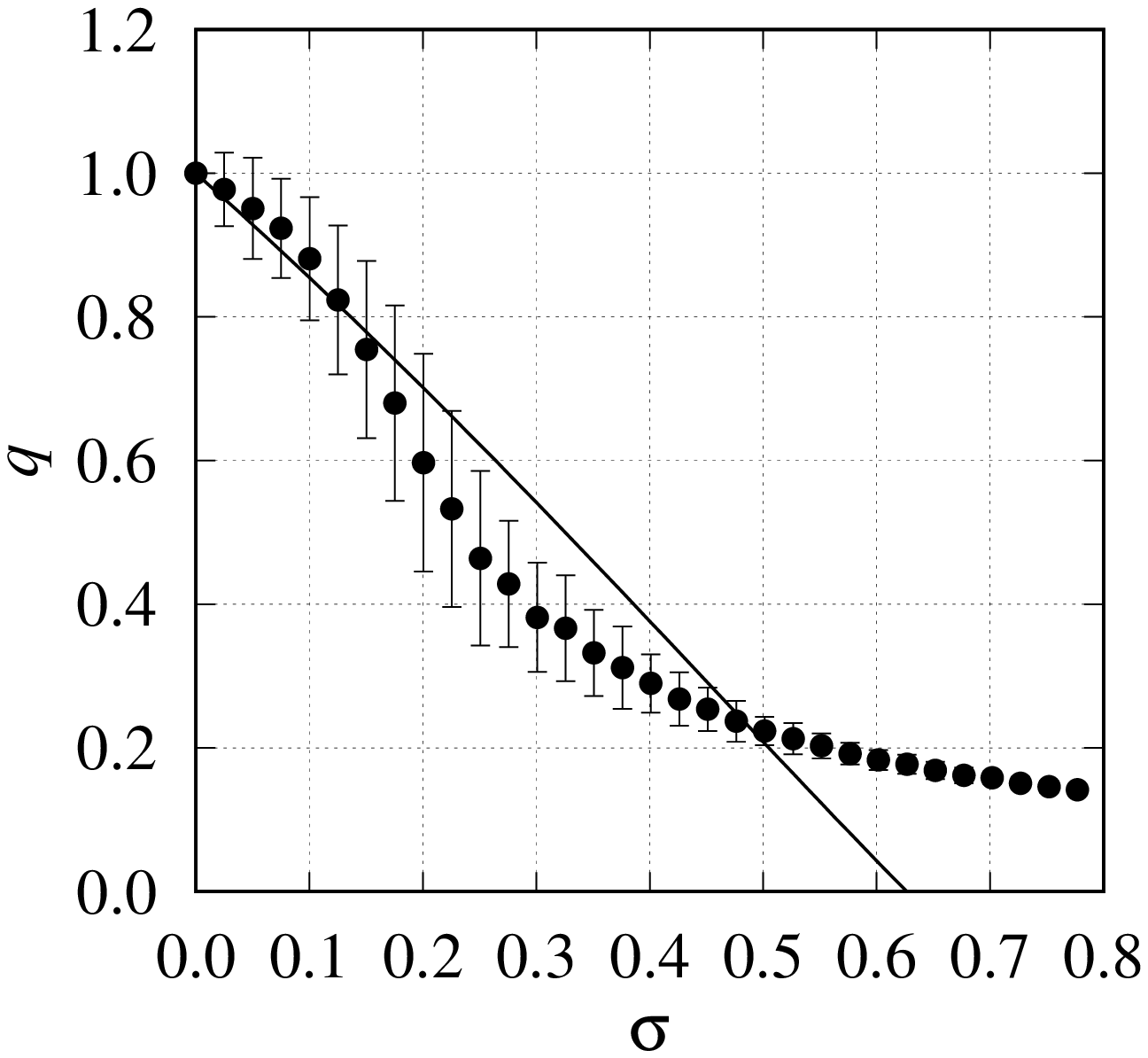} &
   \includegraphics[width=4.0cm]{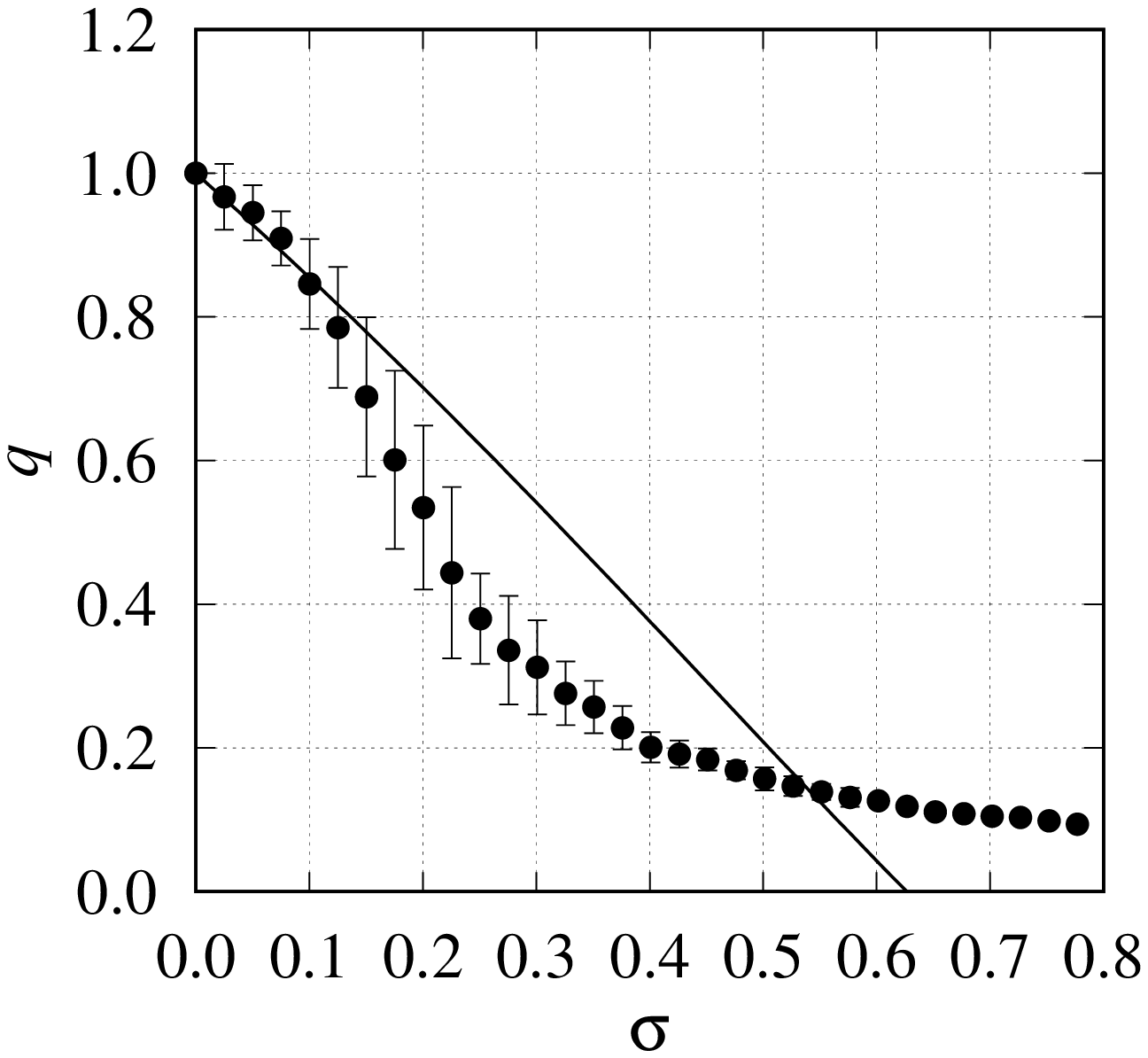}
     \end{tabular}
  \caption{$\sigma$ dependence of sample average of $q$ in  phase oscillator network.
        (a) $N=100$, (b) $N=200$, (c) $N=500$.}
%    (a) $N=100$, (b) $N=500$.}
 \label{fig:OSC_q-sigma}
  \end{center}
\end{figure}
    %fig3
\begin{figure}[H]
  \begin{center}
    \begin{tabular}{ccc}
      (a) & (b) & (c) \\
   \includegraphics[width=4.0cm]{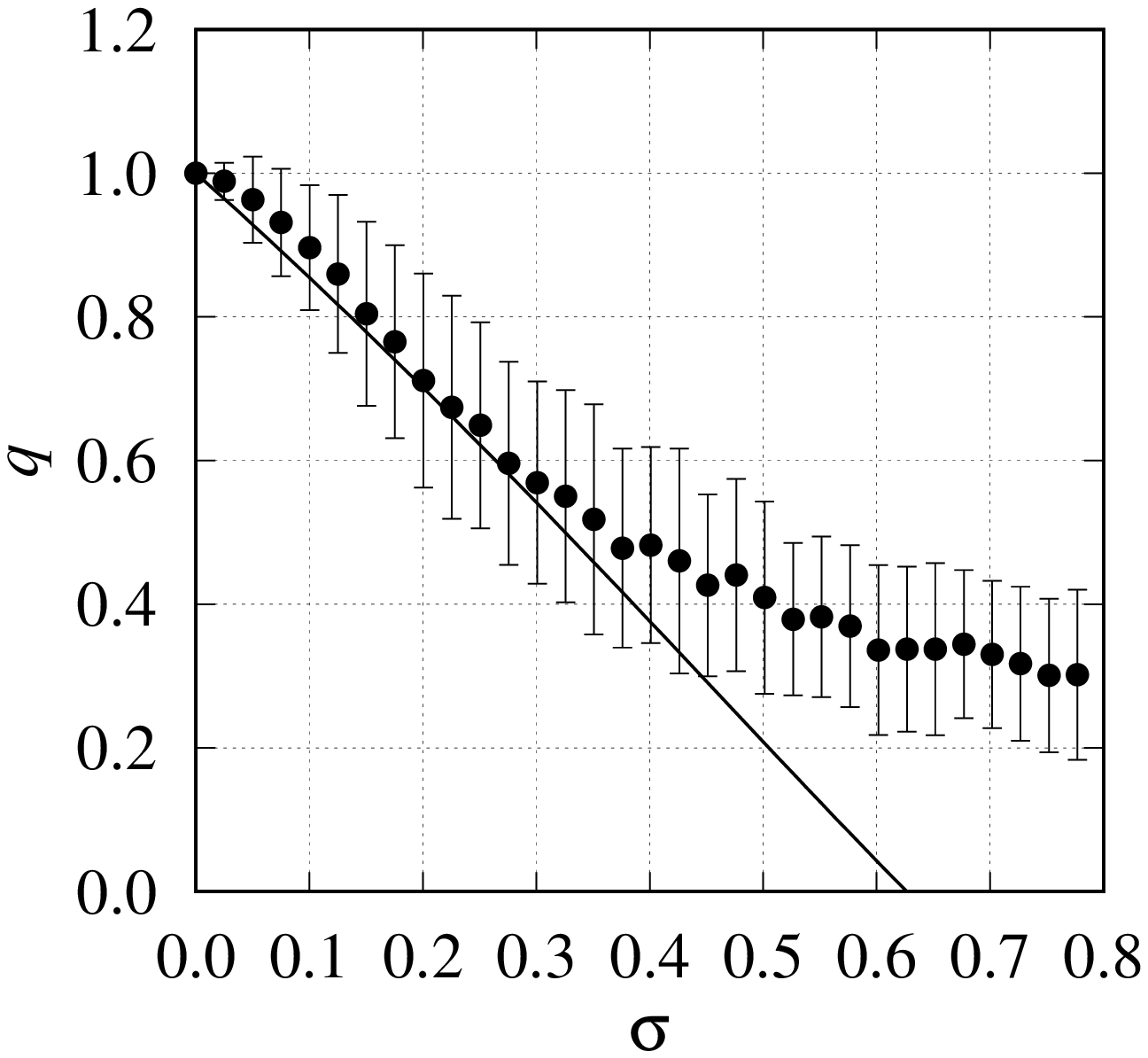} &
   \includegraphics[width=4.0cm]{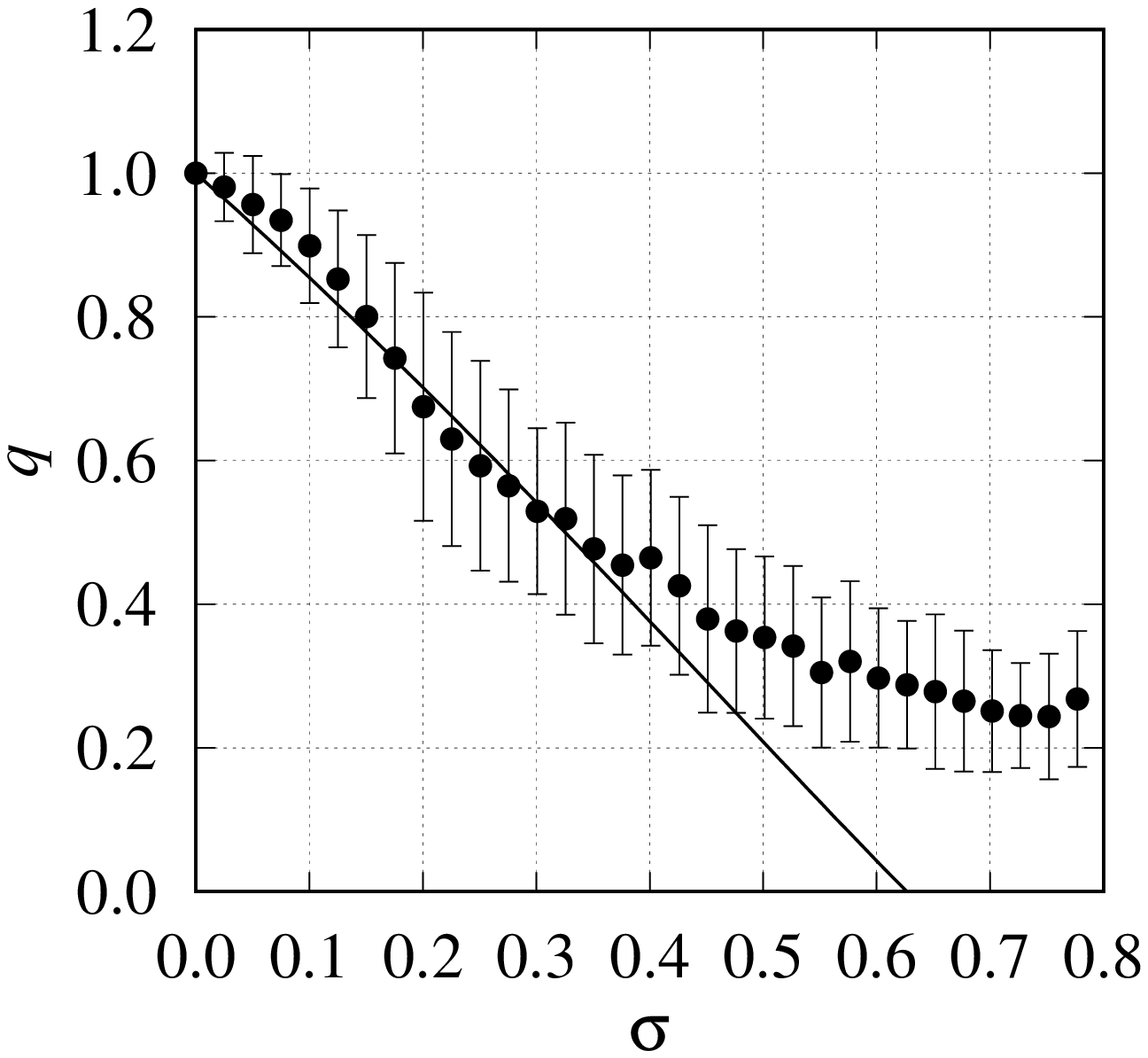} &
   \includegraphics[width=4.0cm]{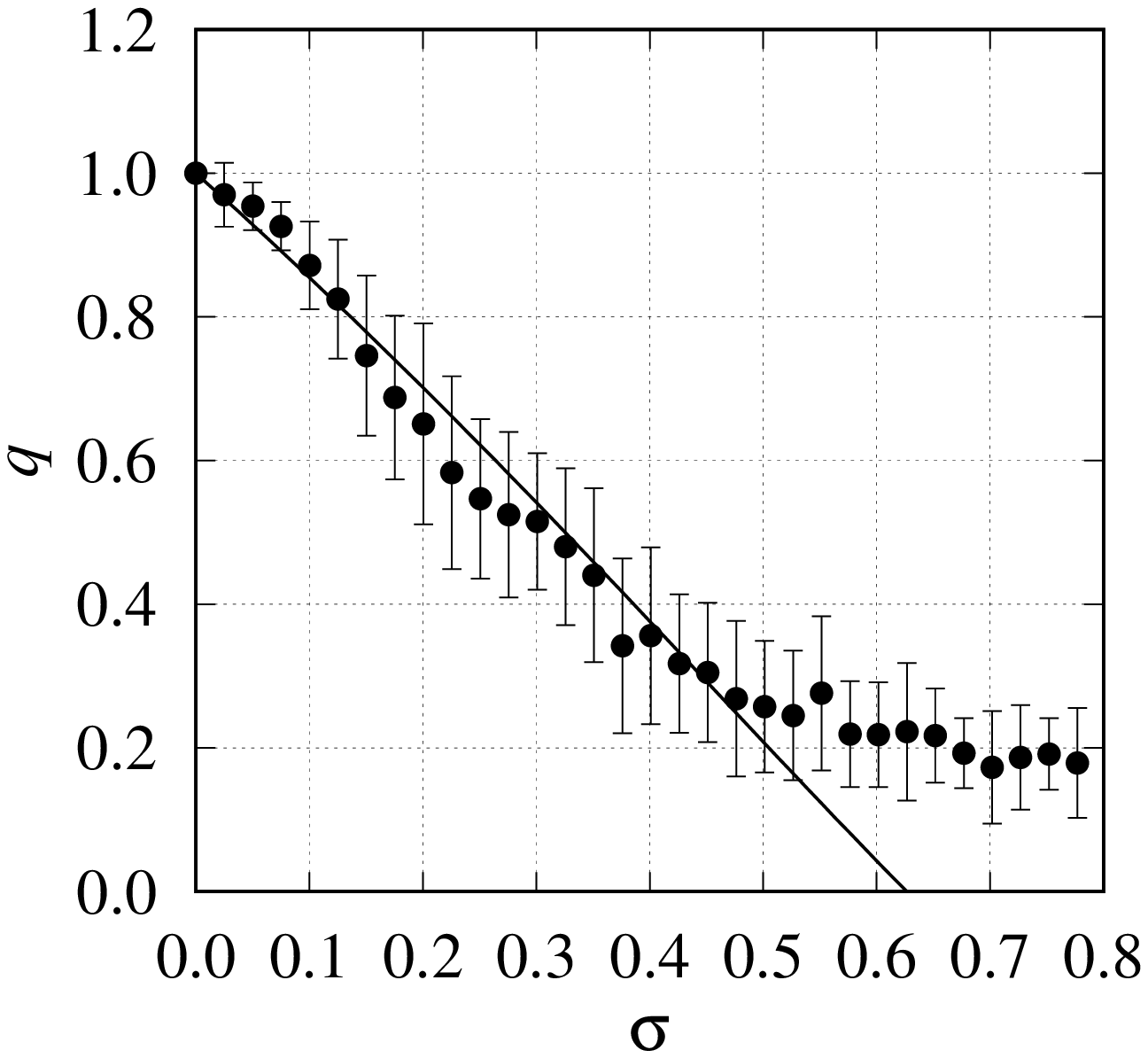} 
     \end{tabular}
  \caption{$\sigma$ dependence of sample average of $q_{\rm av}$ in  phase oscillator network.
    (a) $N=100$, (b) $N=200$, (c) $N=500$.}
%       (a) $N=100$, (b) $N=500$.}    
 \label{fig:OSC_qav-sigma}
  \end{center}
\end{figure}
The results of $T$ dependences of $q$ in the XY 
 model and $\sigma$ dependences of $q$ in the phase oscillator network 
 imply that they differ by the factor $\sqrt{\pi/2}$ in the scale of abscissa axes
  as expected.\par
  \subsection{ Local Field}
  Now, let us study the local field $p_j=x_j + i \; y_j$
 which is defined by 
\begin{equation}
  %  p_j = \frac{1}{N} \sum_{k=1}^N J_{jk} \; e^{i \phi_k}.  \label{eq:LF}
    p_j = \sum_{k=1}^N J_{jk} \; e^{i \phi_k}.  \label{eq:LF}
\end{equation}
 LFs move on the complex plane 
 with  time due to the thermal fluctuation in the XY model, 
 and in the phase oscillator network they move on the complex plane with time
  according to the evolution equation (\ref{eq:TimeUpdate}). 
\subsubsection{XY model}
%  Firstly, in the XY model,
We numerically examined the  spatial distribution of
LFs on the complex plane for all spins.
    The initial values of $\phi_j \ (1\le j \le N)$ were set as the final equilibrium state
    obtained when we calculated $q$.
    In Fig. \ref{fig:XY_LFbunpu}, we display the distribution of LFs on the complex plane 
    and the  probability density $P(r)$ of LFs, where $r=\sqrt{x^2+y^2}$.
    To draw Fig.  \ref{fig:XY_LFbunpu}, a Monte Carlo simulation was carried out 
    for $N=500$ and data  were taken every 1 MC sweep during 10000 MC sweeps.
   That is, $10000 \times N$ data are used   to draw  Fig.  \ref{fig:XY_LFbunpu}.
When $ T $ is low, $P(r)$ is a volcanic shape with a hole in the center, {\it i.e.,} $r=0$, 
and the hole gradually closes with the increase of $T$, and then
it disappears and the peak position becomes $r=0$ for $T>0.5 (=T_c)$.
%fig4
\begin{figure}[H]
 \begin{center}
   \begin{tabular}{cc}
(a) & (b) \\     
	\includegraphics[width=4.2cm]{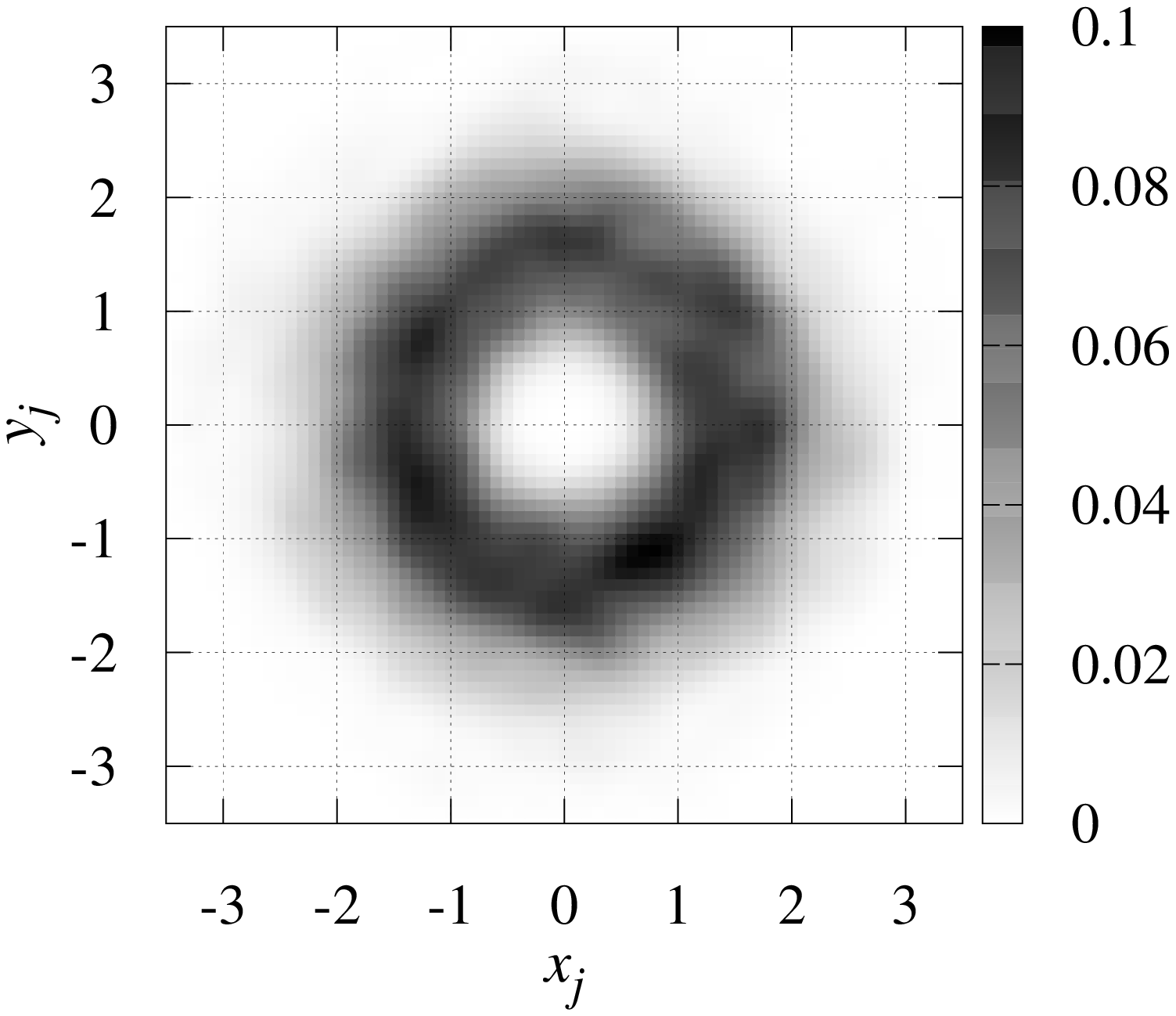} &
	\includegraphics[width=4.0cm]{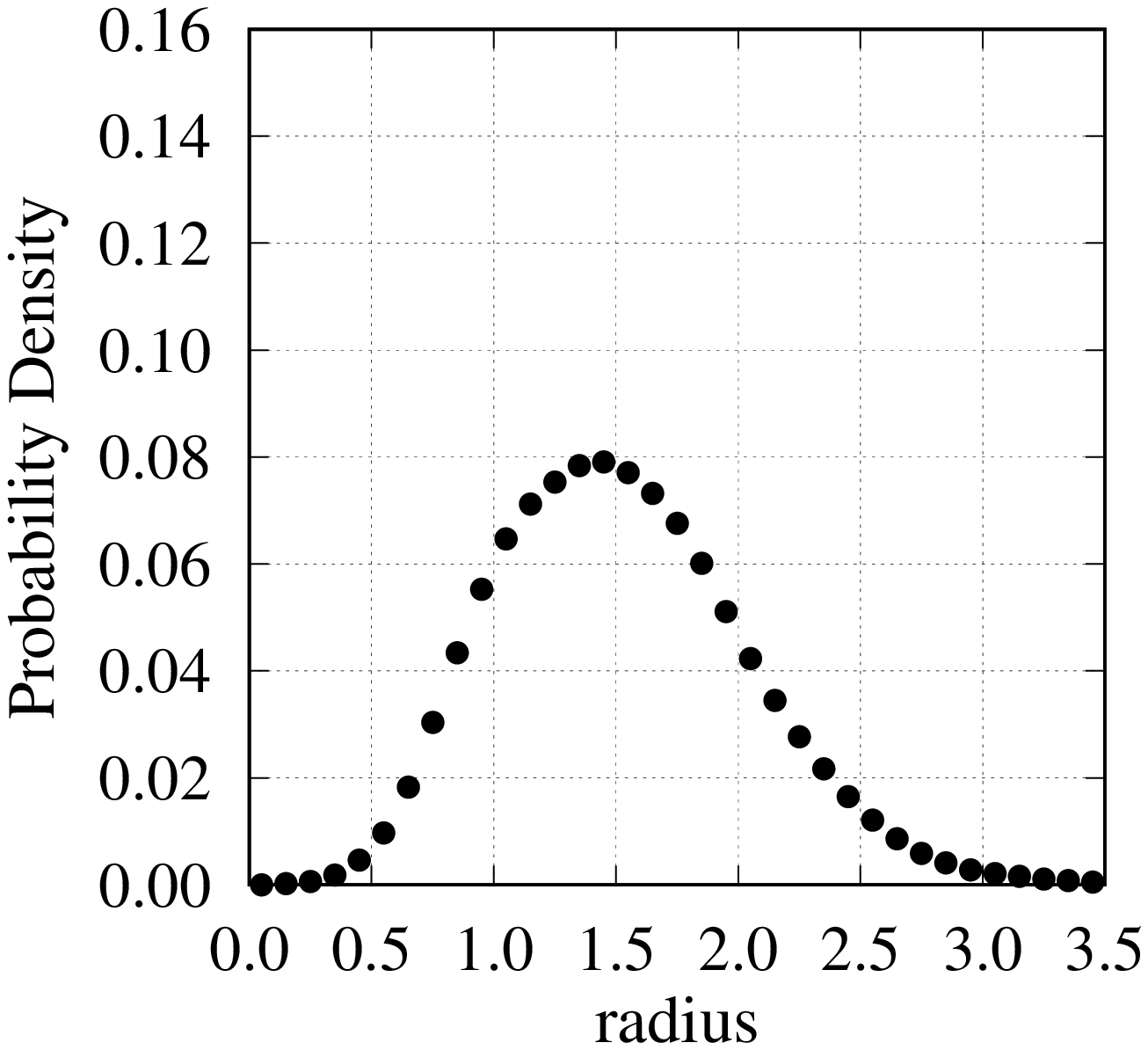} \\
%	\includegraphics[width=4.2cm]{SK-XY_Prob(R)_T=0.04-map-2.eps} &
%	\includegraphics[width=4.0cm]{SK-XY_Prob(R)_T=0.04-map-3.eps} \\ 
        %	\small{(a) $T=0.04$} & \small{(b) $T=0.04$} \\
        (c) & (d) \\     
	\includegraphics[width=4.2cm]{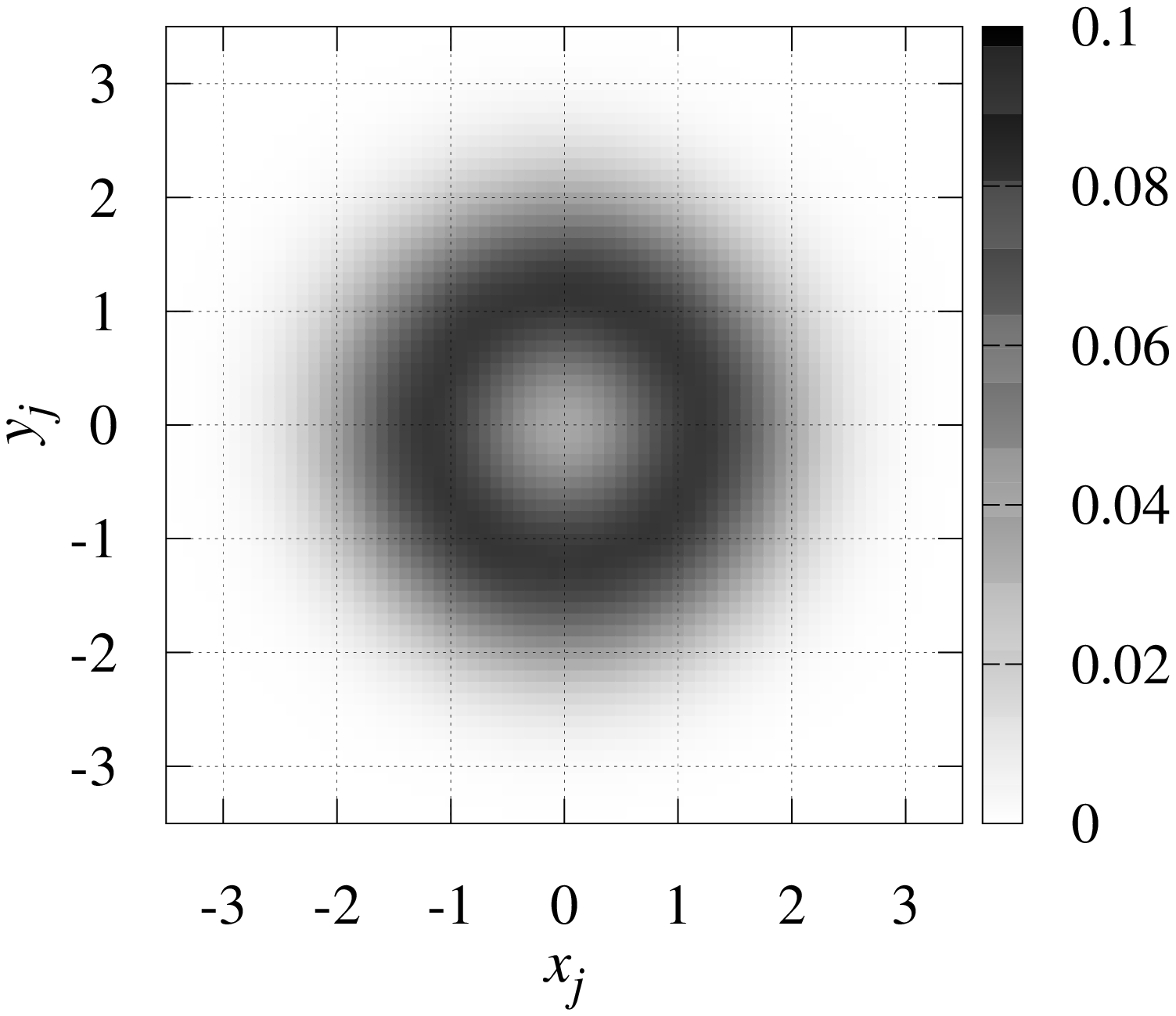}&
	\includegraphics[width=4.0cm]{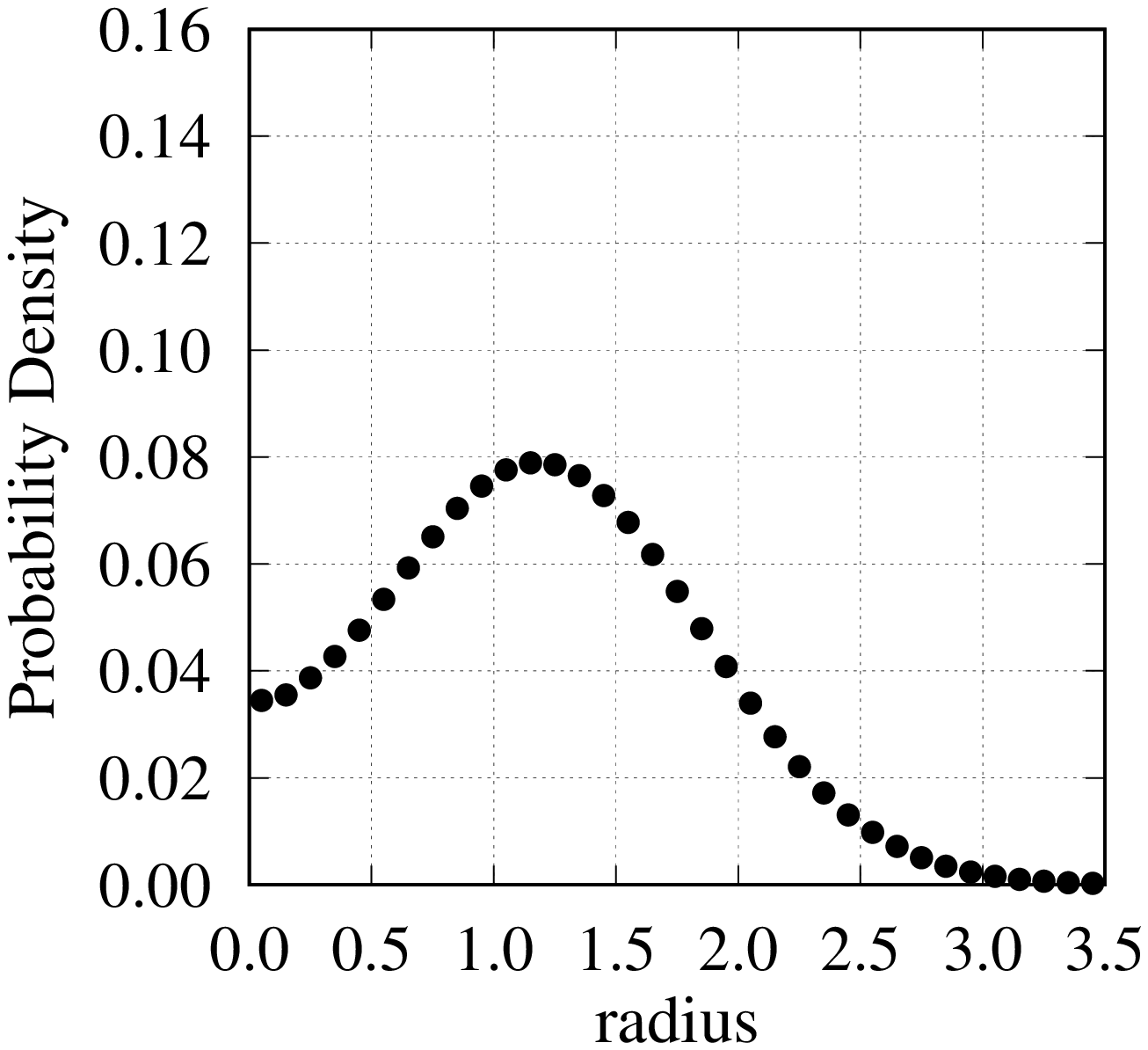}\\
%	\includegraphics[width=4.2cm]{SK-XY_Prob(R)_T=0.26-map-2.eps} & 
%	\includegraphics[width=4.0cm]{SK-XY_Prob(R)_T=0.26-map-3.eps} \\
        %	\small{(c) $T=0.26$} & \small{(d) $T=0.26$} \\
                (e) & (f) \\     
	\includegraphics[width=4.2cm]{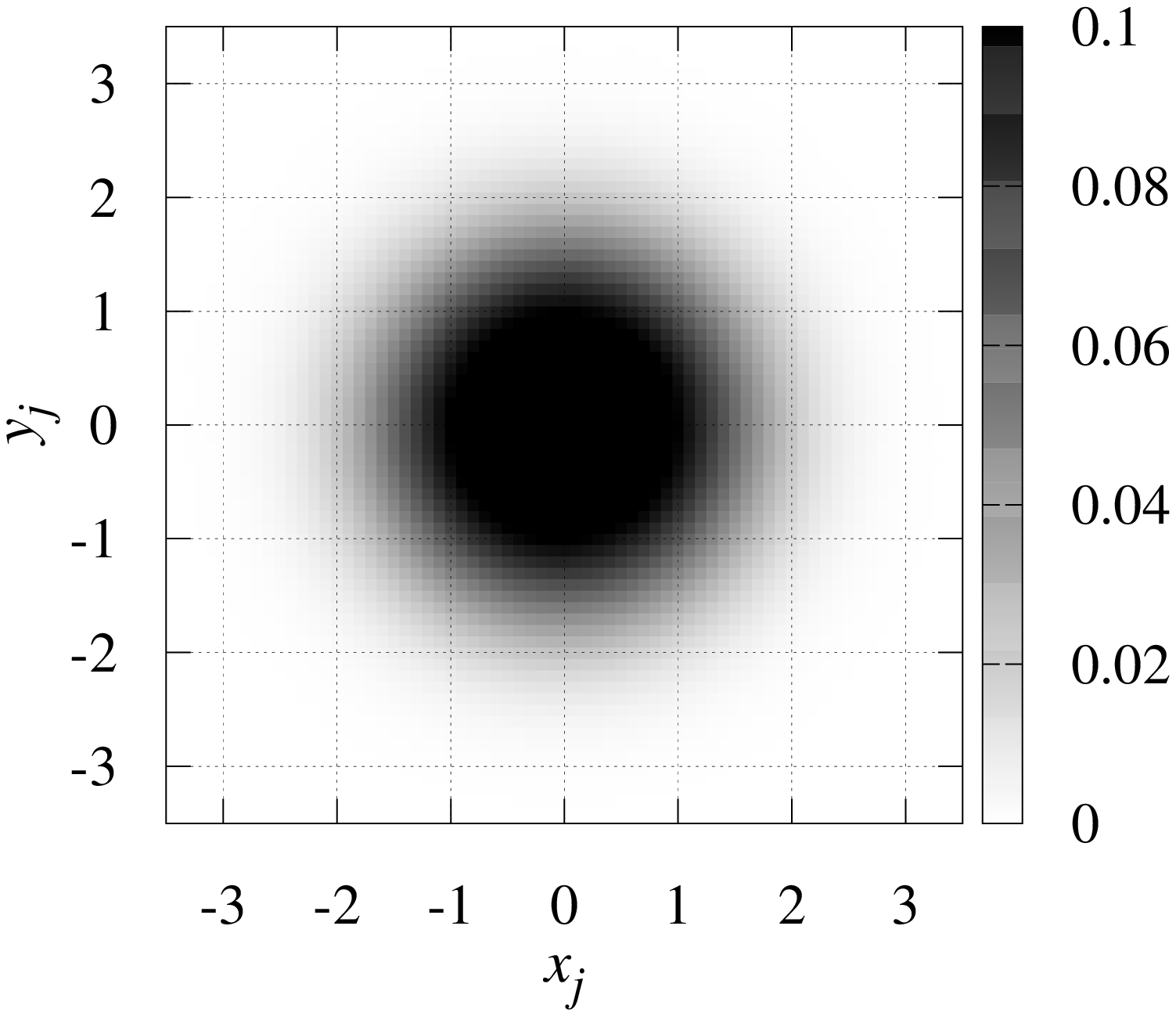}& 
	\includegraphics[width=4.0cm]{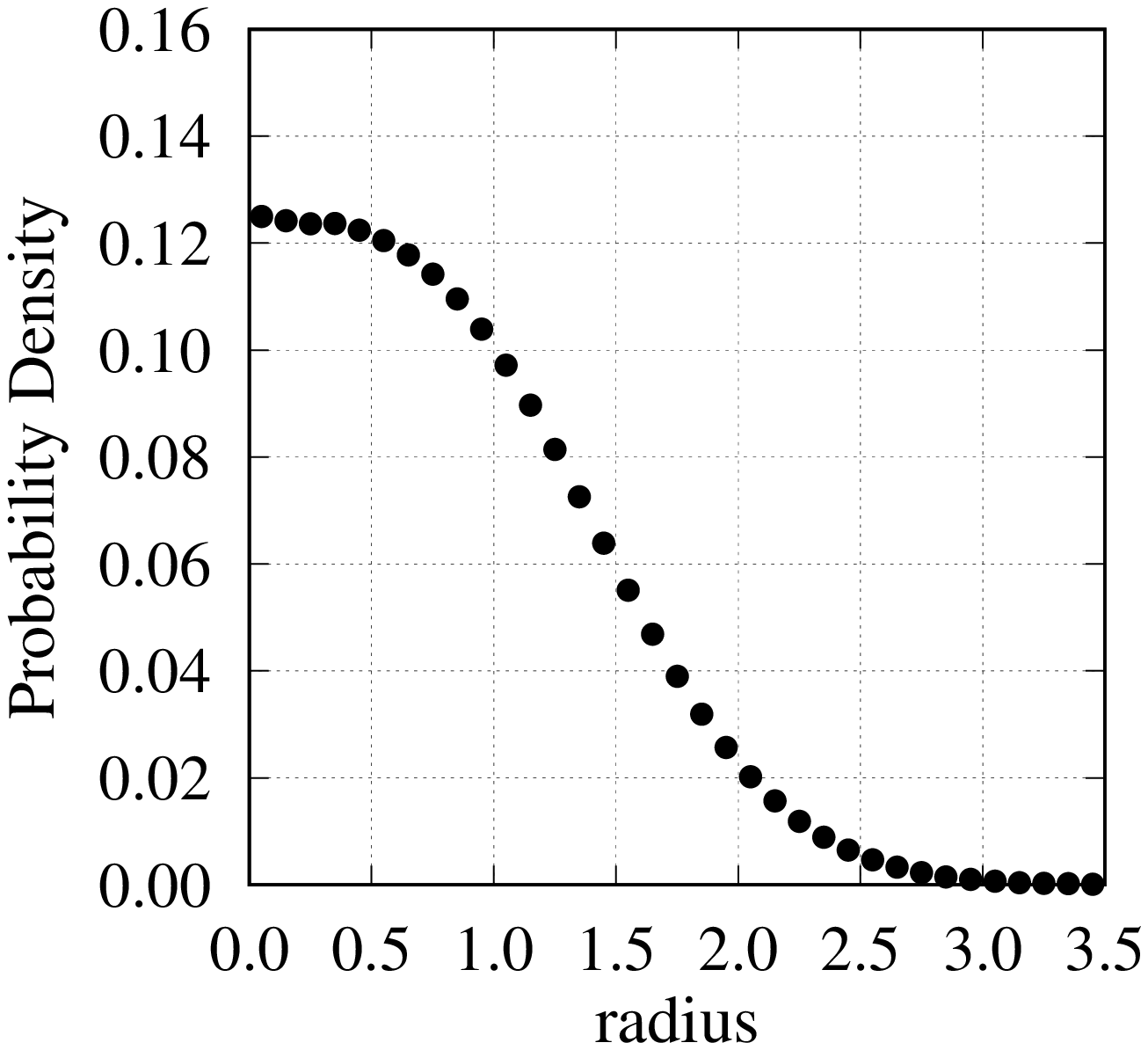}\\
%	\includegraphics[width=4.2cm]{SK-XY_Prob(R)_T=0.50-map-2.eps} & 
%	\includegraphics[width=4.0cm]{SK-XY_Prob(R)_T=0.50-map-3.eps} \\
        %	\small{(e) $T=0.5$} & \small{(f) $T=0.5$} \\
                (g) & (h) \\             
	\includegraphics[width=4.2cm]{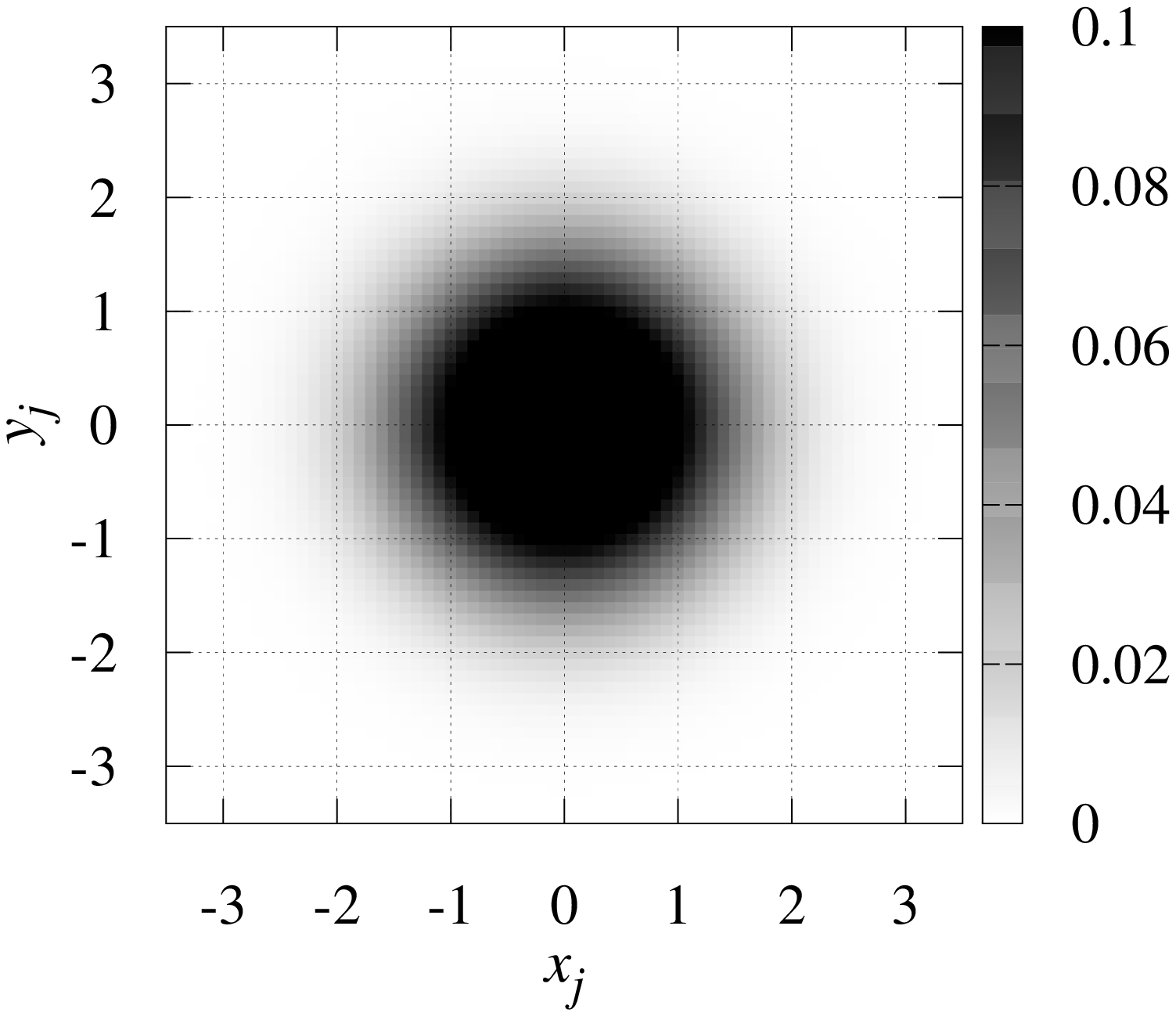}&
	\includegraphics[width=4.0cm]{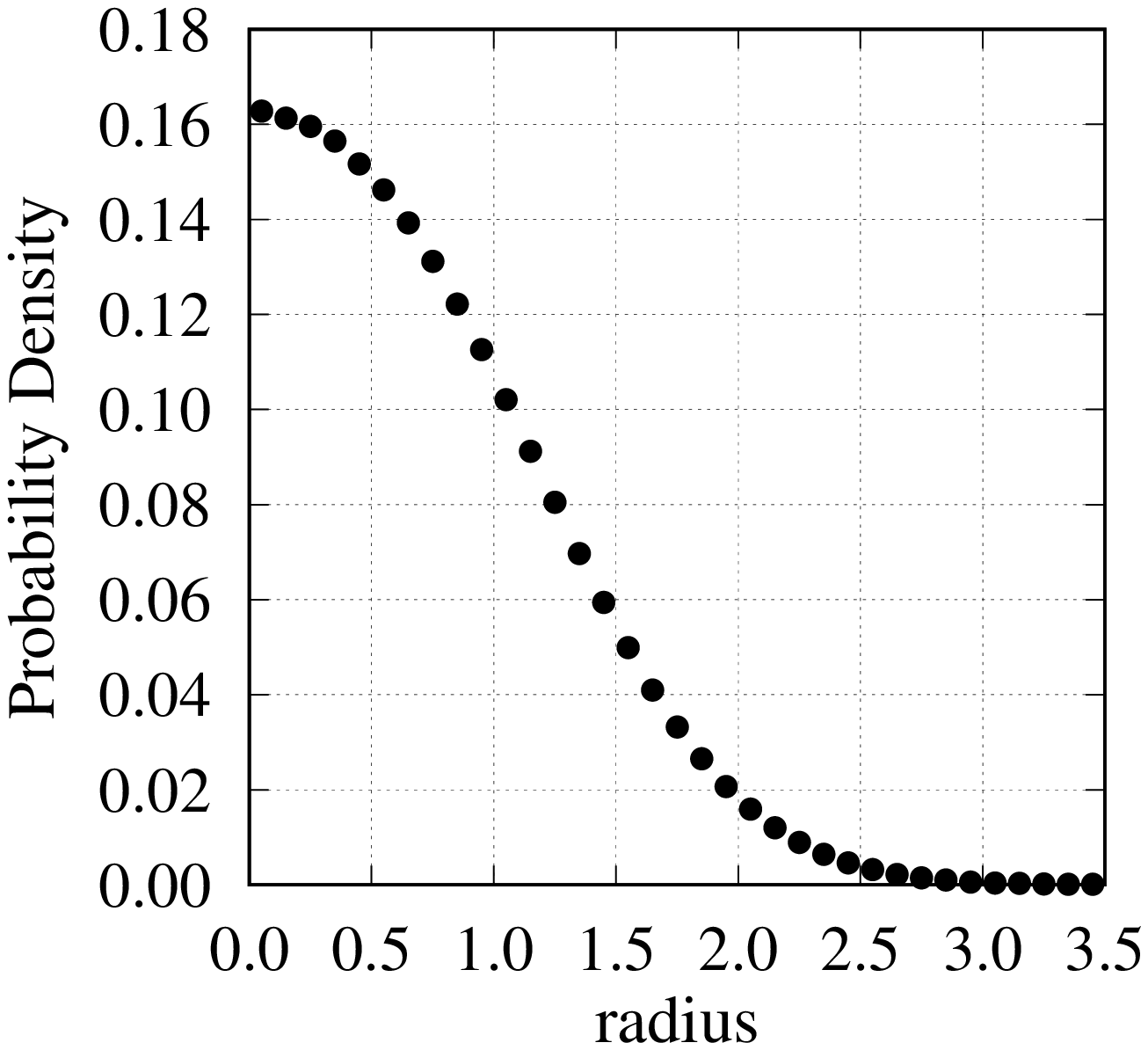}
%	\includegraphics[width=4.2cm]{SK-XY_Prob(R)_T=0.60-map-2.eps} & 
%	\includegraphics[width=4.0cm]{SK-XY_Prob(R)_T=0.60-map-3.eps} \\
%	\small{(g) $T=0.6$} & \small{(h) $T=0.6$}  
    \end{tabular}
 \vspace{-5mm}
 \caption{Local field of XY model ($N=500$). Left panel:  spatial distribution of LFs
    on the complex plane.
    Right panel: probability density of LFs, $P(r)$.
  (a), (b) $T=0.04$, (c), (d) $T=0.26$, (e), (f)  $T=0.5$, (g), (h) $T=0.6$. }
 \label{fig:XY_LFbunpu}
  \end{center}
\end{figure}
\subsubsection{Phase oscillator network}
We calculated LFs as in the XY model.
The initial values of $\phi_j \ (1\le j \le N)$ were set as the final  state 
   obtained when we calculated $q$. 
In Fig. \ref{fig:OSC_LFbunpu}, we display the distribution of LFs and $P(r)$.
 A  computer simulation was carried out for  $N=500$ until $t=10000$,
 and data were taken  every  time interval 1
 to draw Fig. \ref{fig:OSC_LFbunpu}.
 That is, the number of data to draw Fig. \ref{fig:OSC_LFbunpu} is
 % one-fifth of that in the XY model.
  the same as  in the XY model. 
 As is seen from Fig. \ref{fig:OSC_LFbunpu}, with the increase of $\sigma$
  from 0,
  behavior of  $P(r)$ is the same as in the XY model  
 and the peak position becomes $r=0$
 for $\sigma>0.5 \sqrt{\pi/2}(=T_c \sqrt{\pi/2})$.
 % \end{document}
%fig5 
\begin{figure}[H]
 \begin{center}
   \begin{tabular}{cc}
     (a) & (b)\\
	\includegraphics[width=4.2cm]{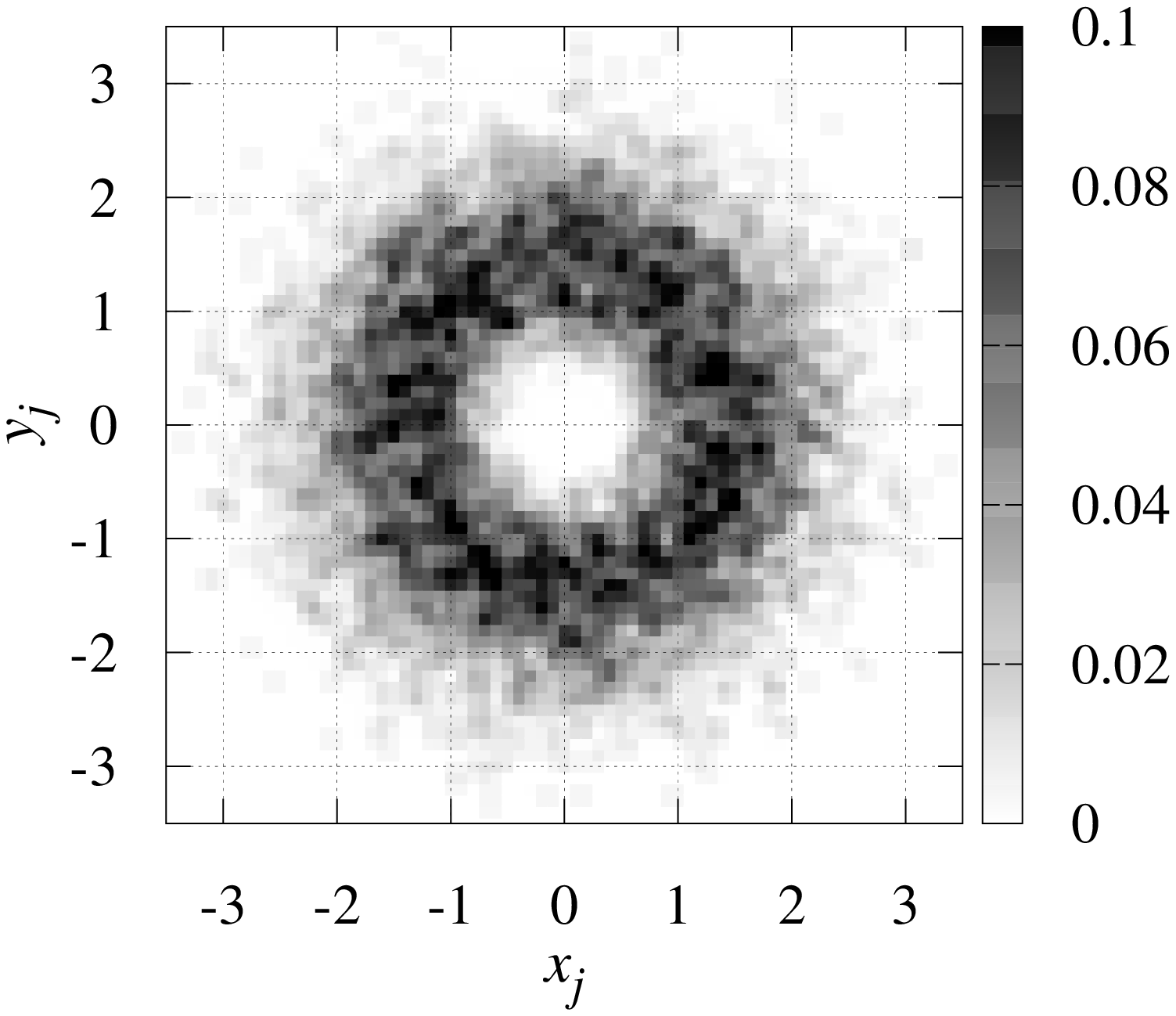} &
	\includegraphics[width=4.0cm]{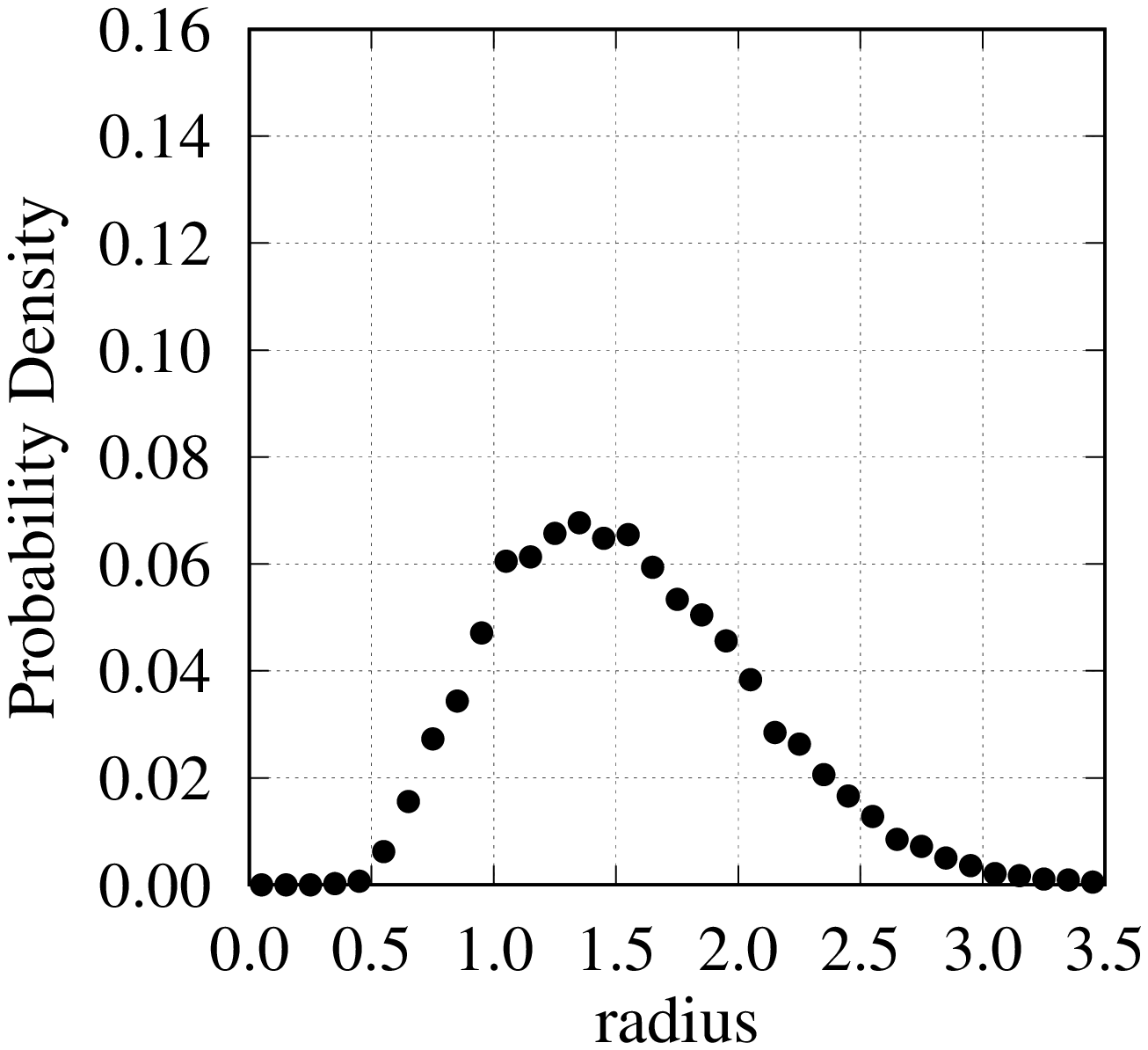} \\
%	\includegraphics[width=4.2cm]{SK-OSC_Prob(R)_sigma=0.04sqrt(0.5pi)-map-2.eps} & 
%	\includegraphics[width=4.0cm]{SK-OSC_Prob(R)_sigma=0.04sqrt(0.5pi)-map-3.eps} \\ 
        %	\small{(a) $\sigma=0.04\sqrt{\pi/2}$} & \small{(b) $\sigma=0.04\sqrt{\pi/2}$} \\
             (c) & (d)\\
	\includegraphics[width=4.2cm]{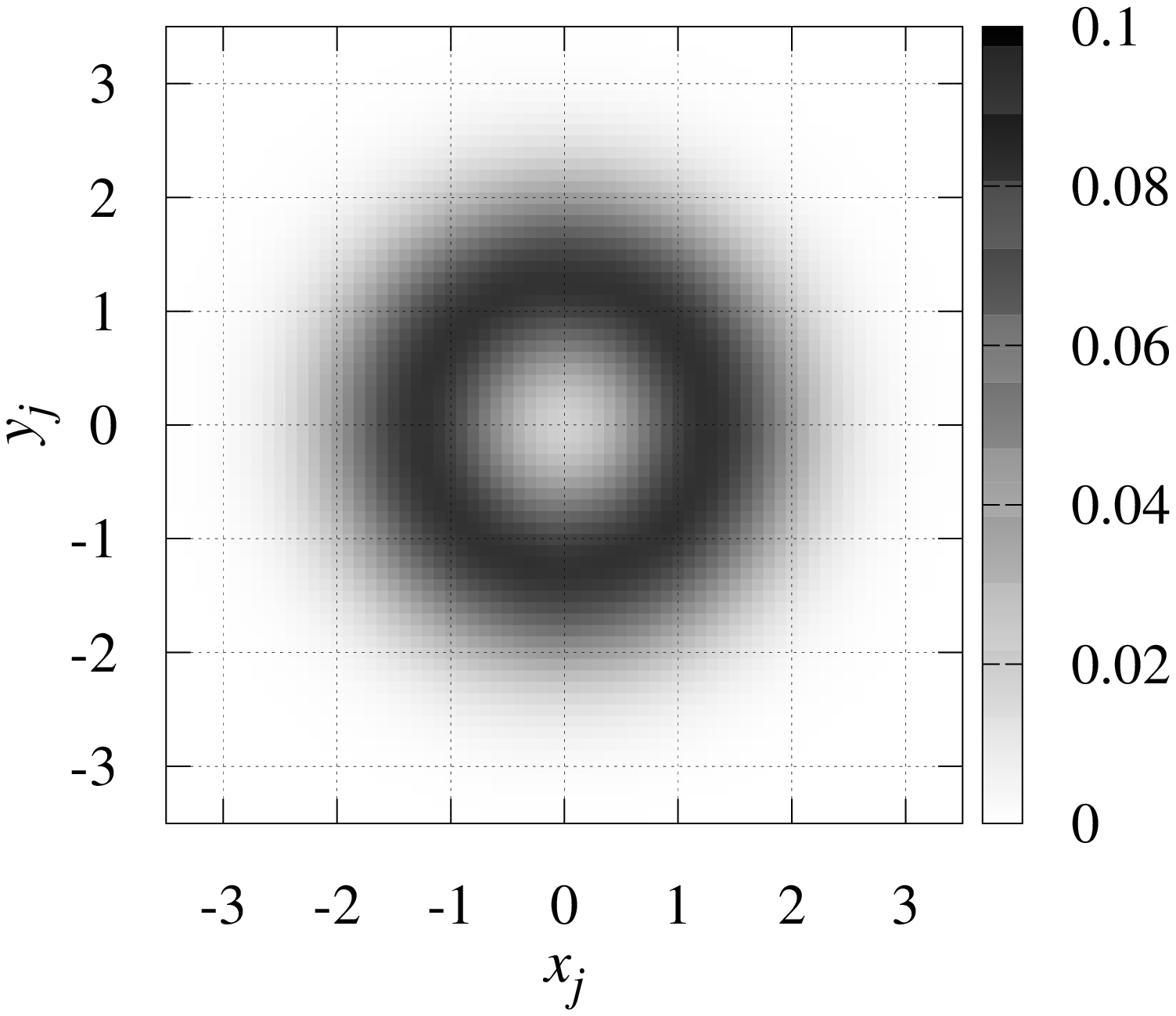} &
	\includegraphics[width=4.0cm]{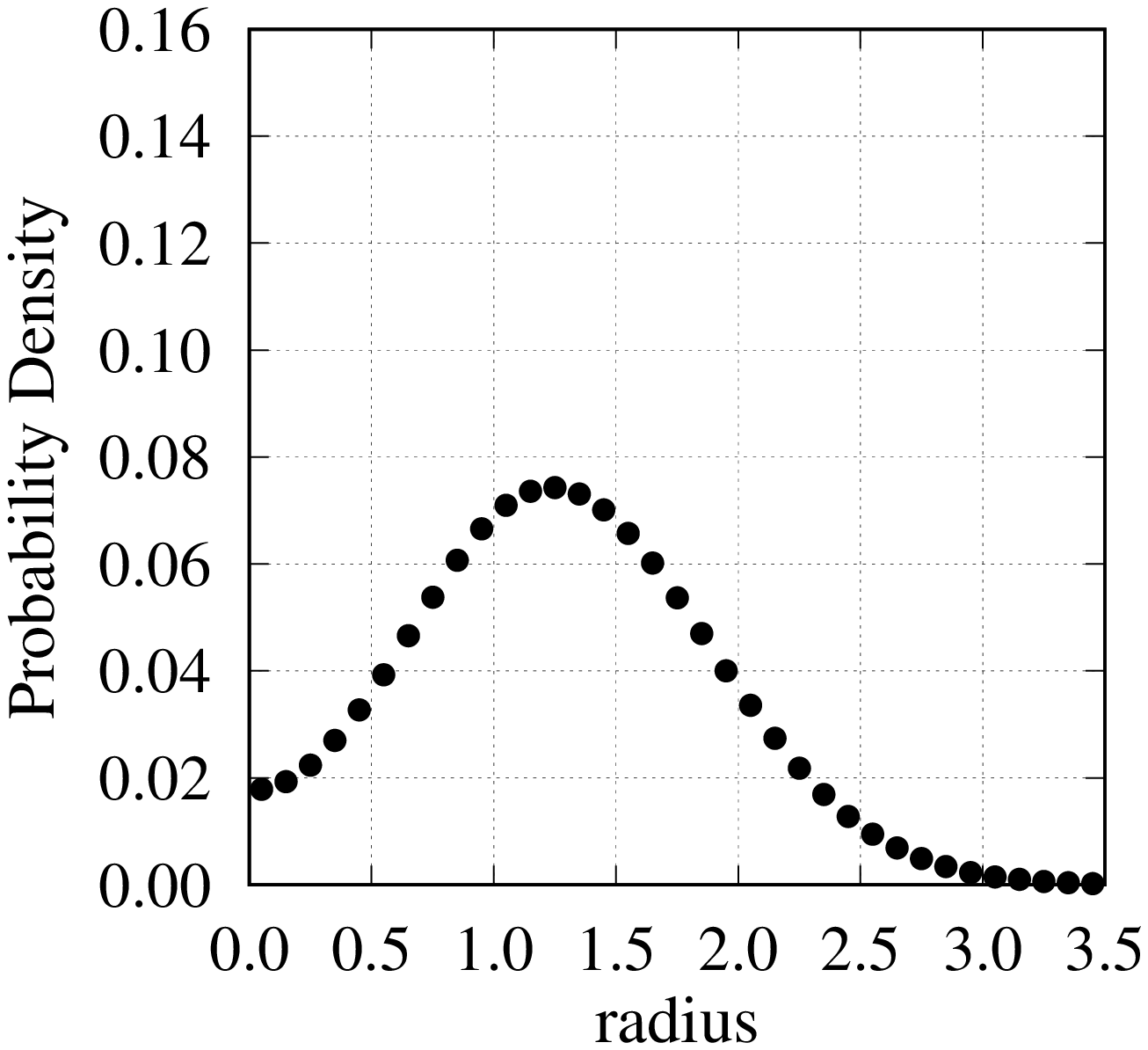} \\
%	\includegraphics[width=4.2cm]{SK-OSC_Prob(R)_sigma=0.26sqrt(0.5pi)-map-2.eps} & 
%	\includegraphics[width=4.0cm]{SK-OSC_Prob(R)_sigma=0.26sqrt(0.5pi)-map-3.eps} \\
        %	\small{(c) $\sigma=0.26\sqrt{\pi/2}$} & \small{(d) $\sigma=0.26\sqrt{\pi/2}$}  \\
               (e) & (f) \\        
	\includegraphics[width=4.2cm]{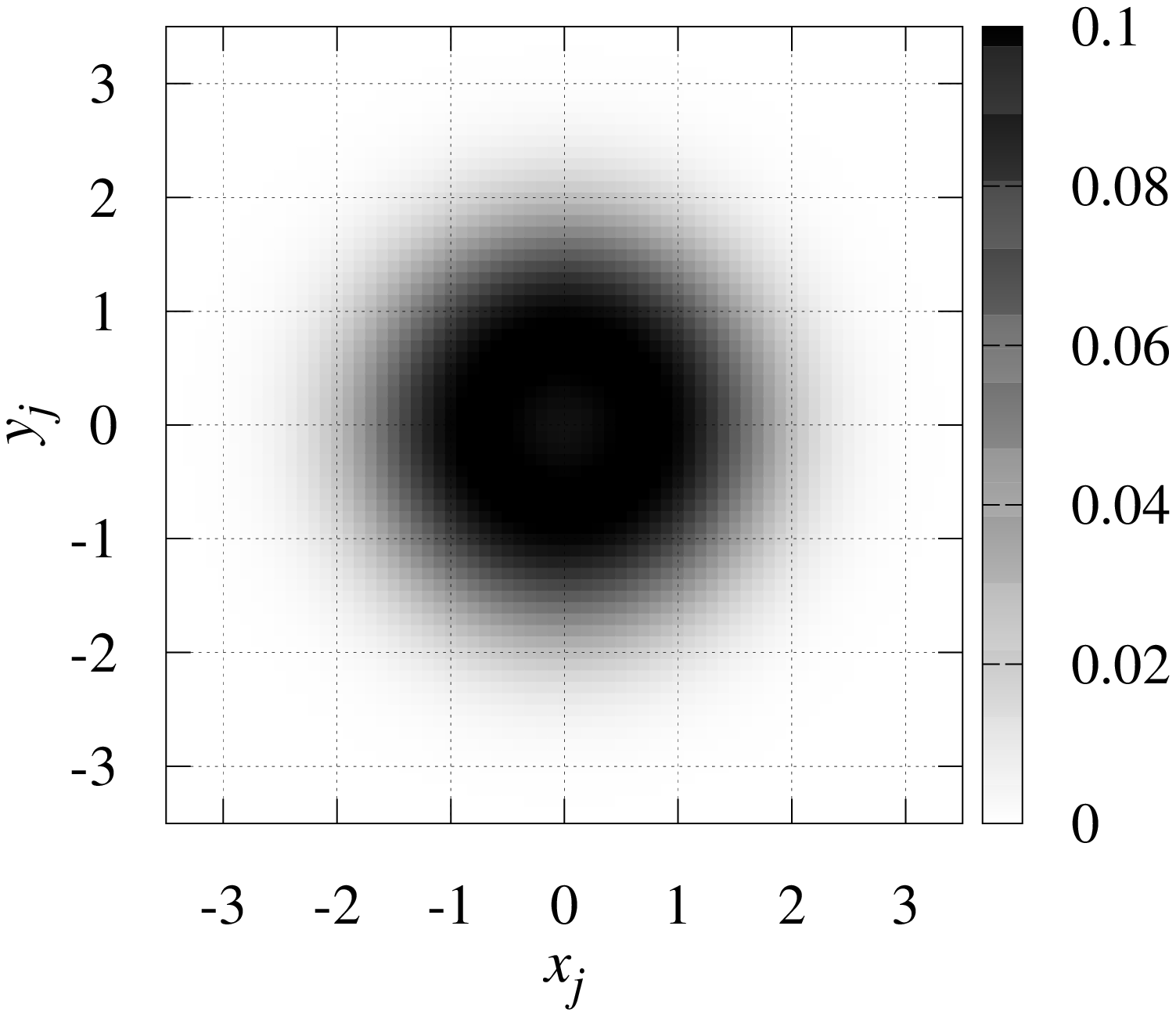} & 
	\includegraphics[width=4.0cm]{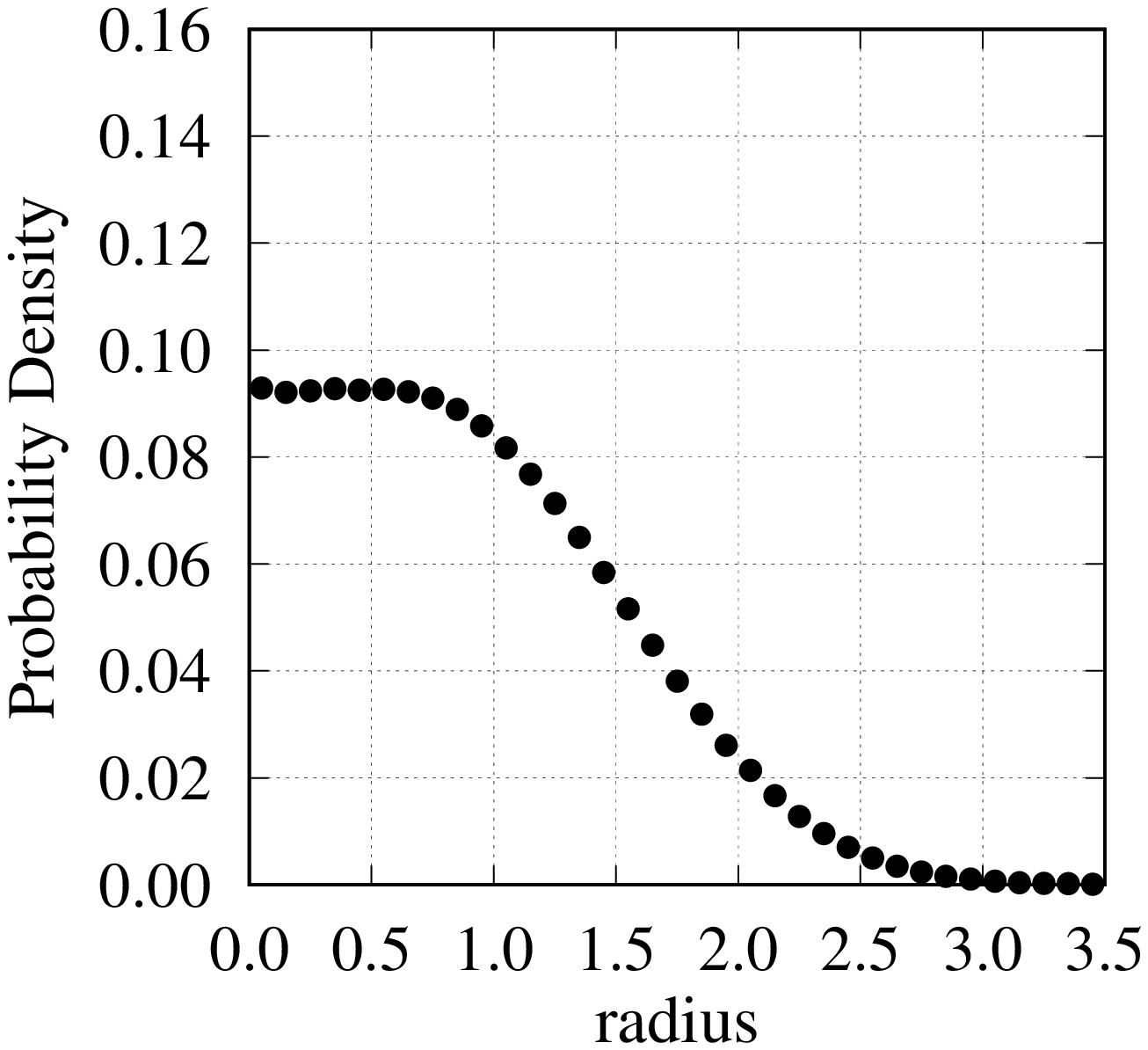} \\
%	\includegraphics[width=4.2cm]{SK-OSC_Prob(R)_sigma=0.50sqrt(0.5pi)-map-2.eps} & 
%	\includegraphics[width=4.0cm]{SK-OSC_Prob(R)_sigma=0.50sqrt(0.5pi)-map-3.eps} \\
        %	\small{(e) $\sigma=0.5\sqrt{\pi/2}$} & \small{(f) $\sigma=0.5\sqrt{\pi/2}$}  \\
                       (g) & (h) \\        
	\includegraphics[width=4.2cm]{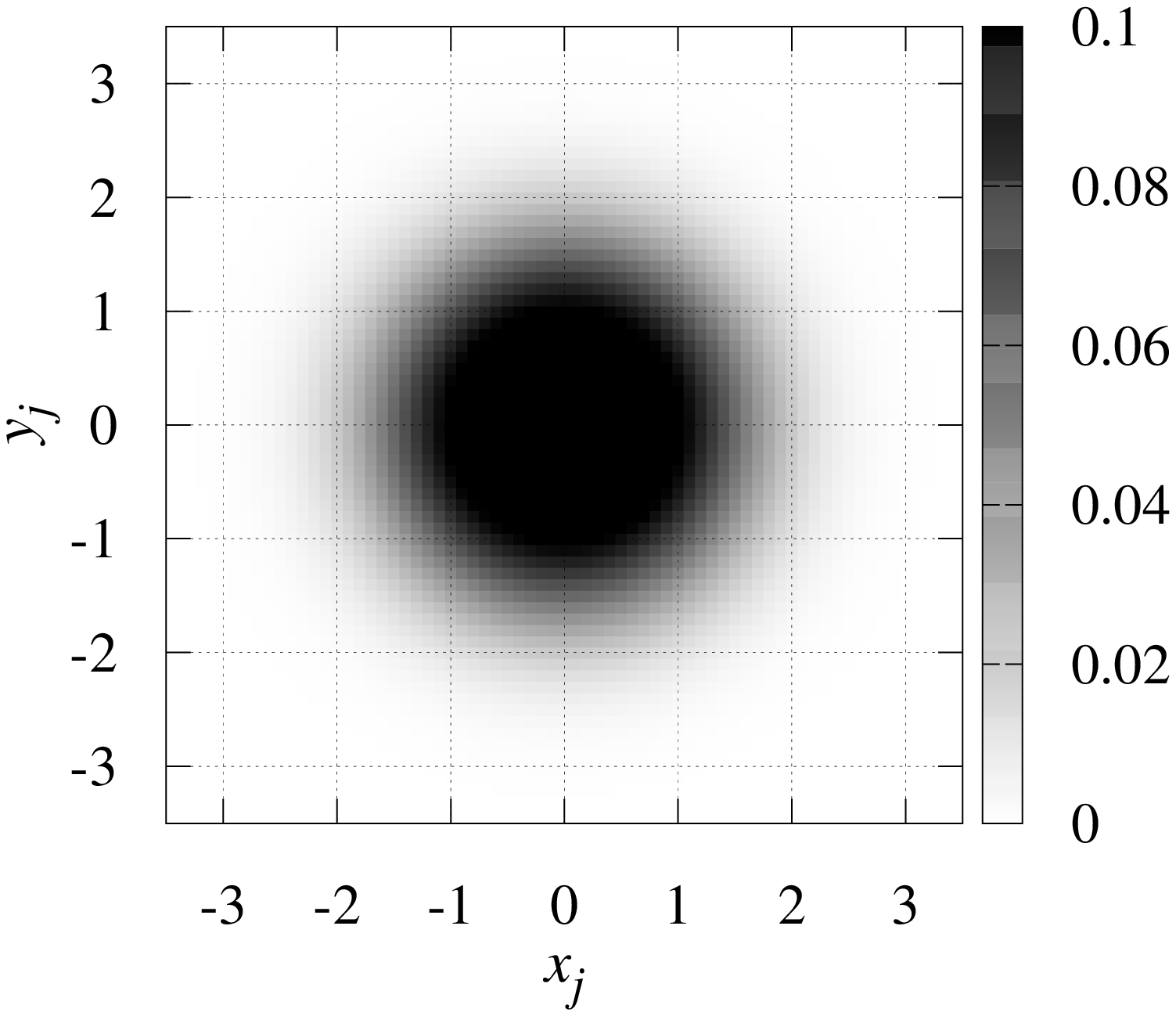} &
	\includegraphics[width=4.0cm]{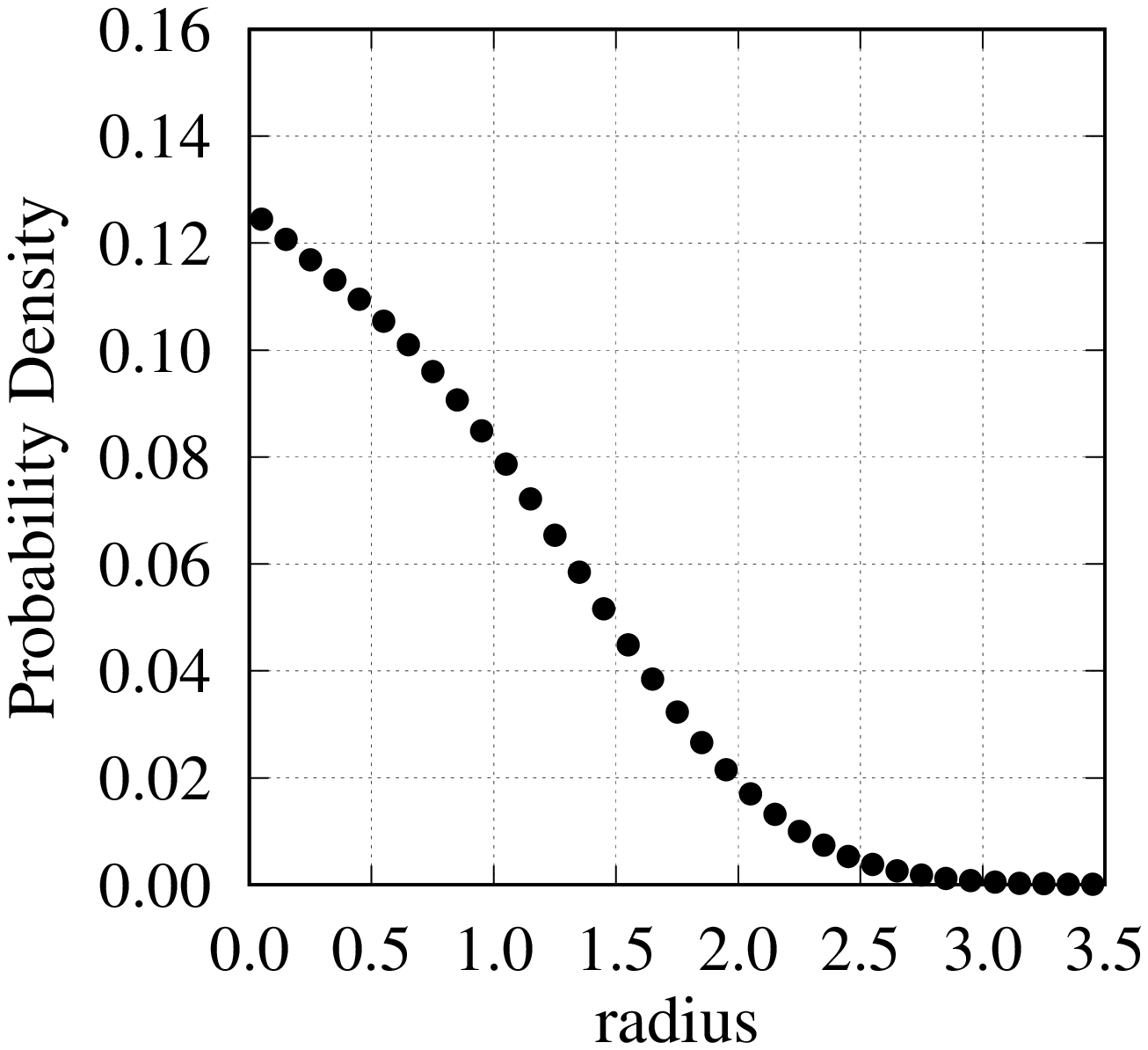} 
%	\includegraphics[width=4.2cm]{SK-OSC_Prob(R)_sigma=0.60sqrt(0.5pi)-map-2.eps} & 
%	\includegraphics[width=4.0cm]{SK-OSC_Prob(R)_sigma=0.60sqrt(0.5pi)-map-3.eps} \\
%	\small{(g) $\sigma=0.6\sqrt{\pi/2}$} & \small{(h) $\sigma=0.6\sqrt{\pi/2}$} \\
    \end{tabular}
 \end{center}
 \vspace{-5mm}
 \caption{   %Local field of phase oscillator network ($N=100$). Left panel:
    Local field of phase oscillator network ($N=500$). Left panel:   
    spatial distribution on the complex plane. Right panel: probability density $P(r)$.
 (a), (b) $T=0.04\sqrt{\pi/2}$, (c), (d) $T=0.26\sqrt{\pi/2}$, (e), (f)  $T=0.5\sqrt{\pi/2}$, (g), (h) $T=0.6\sqrt{\pi/2}$. }
  \label{fig:OSC_LFbunpu}
\end{figure}
\subsubsection{Comparison of results for both models}
In the LFs of the XY model, for $N=500$, 
 the $T$ dependence of  the radius at which the probability density has a peak 
 is shown in Fig. \ref{fig:LF}(a).  We call the radius the peak radius, and denote it by $r_p$.
 The black circles show the peak radius, and the error bars show the radius
 at which the probability density decreases by 5\% from the peak.
 The peak radius at $T>0.5 (=T_c)$ becomes nearly zero.
In the LFs of the phase oscillator network, for $N=500$, 
the $\sigma$ dependence of  the peak radius 
% at which the probability density has a  peak
 is shown in Fig. \ref{fig:LF}(b).
 The circles and error bars have the same meanings as in the XY model.
The peak radius at $\sigma>0.5 \sqrt{\pi/2}(=T_c \sqrt{\pi/2})$ becomes nearly zero.
$T$ and $\sigma$ in which the peak radius becomes zero 
seem to differ by the factor $\sqrt{\pi/2}$ in the scale of abscissa axes
as expected.
%fig6 
\begin{figure}[H]
 \begin{center}
   \begin{tabular}{cc}
     (a) & (b) \\
	\includegraphics[width=4.0cm]{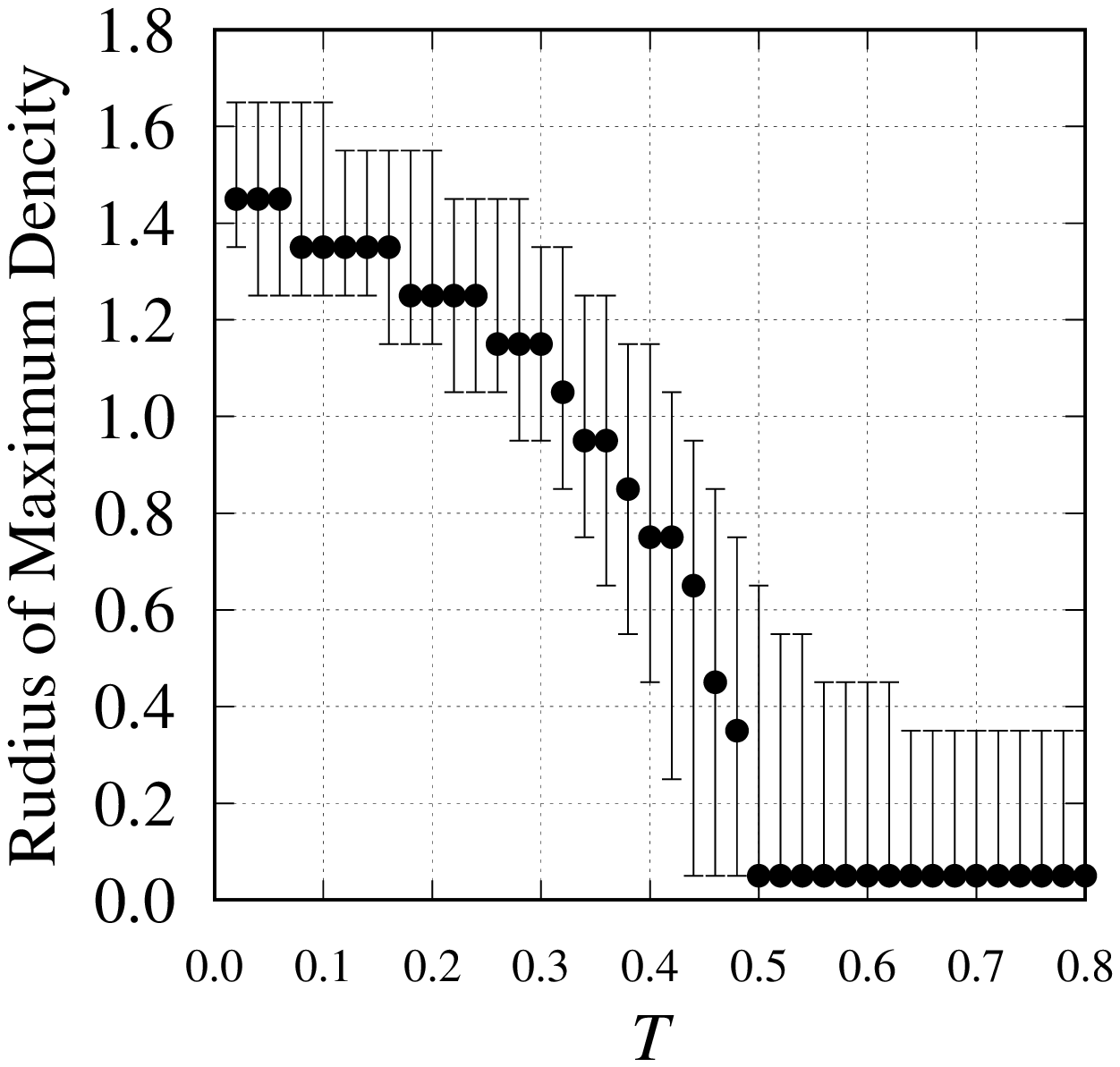} &
	\includegraphics[width=4.0cm]{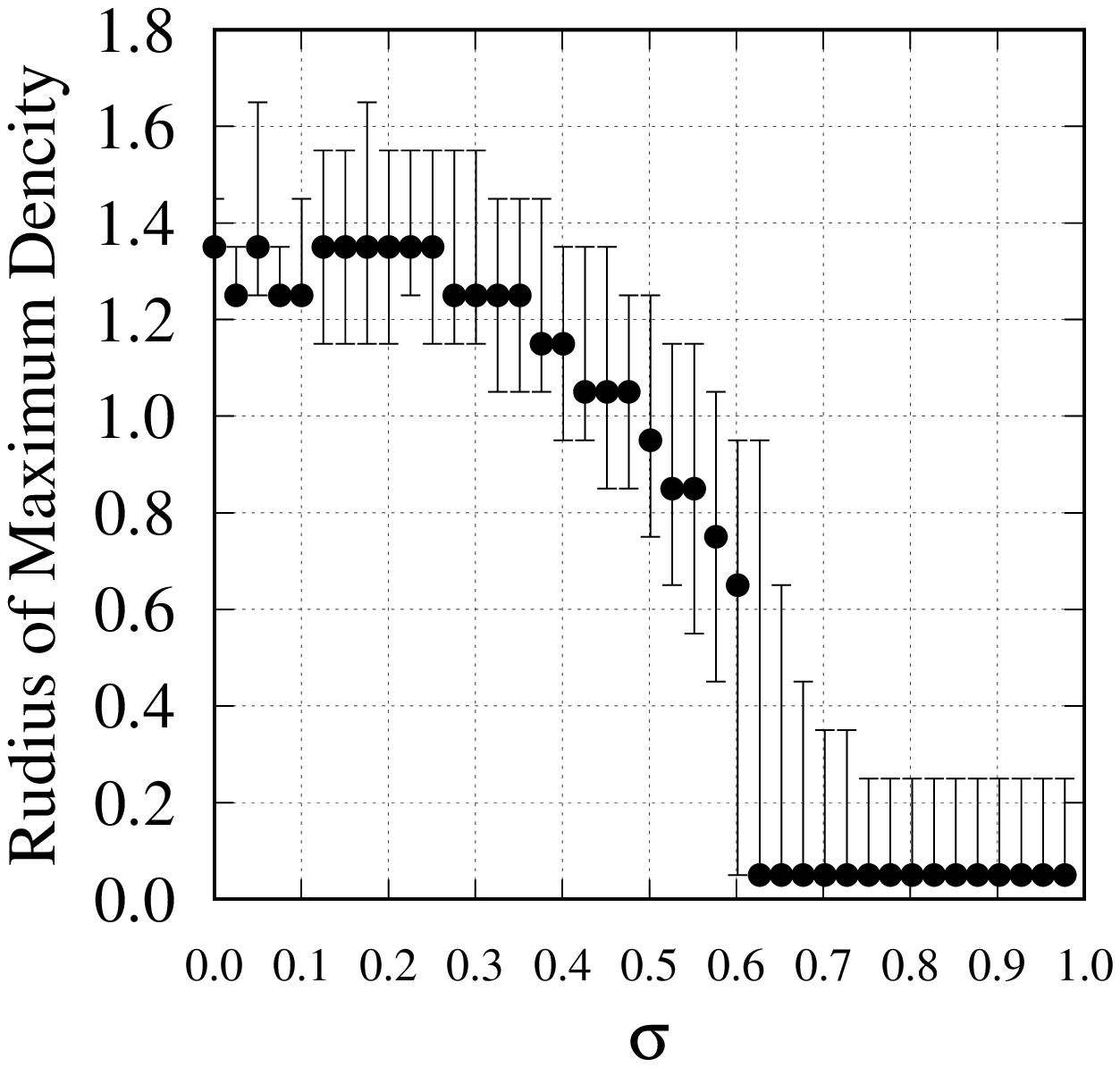}
%	\includegraphics[width=4.0cm]{SK-XY_peakProb(R)_T-4.eps}  &
%	\includegraphics[width=4.0cm]{SK-OSC_peakProb(R)_sigma-4.eps} \\
%	\small(a) XY model &   \small(b) phase oscillator network
    \end{tabular}
	\vspace{-3mm}
        \caption{Temperature dependence of the peak radius $r_p$          %position of $P(r)$ 
  in the XY model ($N=500$) 
  and $\sigma$ dependence of $r_p$ in  the phase oscillator network ($N=500$).
        (a) XY model, (b) phase oscillator}
  \label{fig:LF}
 \end{center}
\end{figure}
%fig7 
\begin{figure}[H]
 \begin{center}
   \begin{tabular}{c}
     %     \small(a)  &   \small(b)  &(c) \\
     (a) \\
\includegraphics[width=5.0cm]{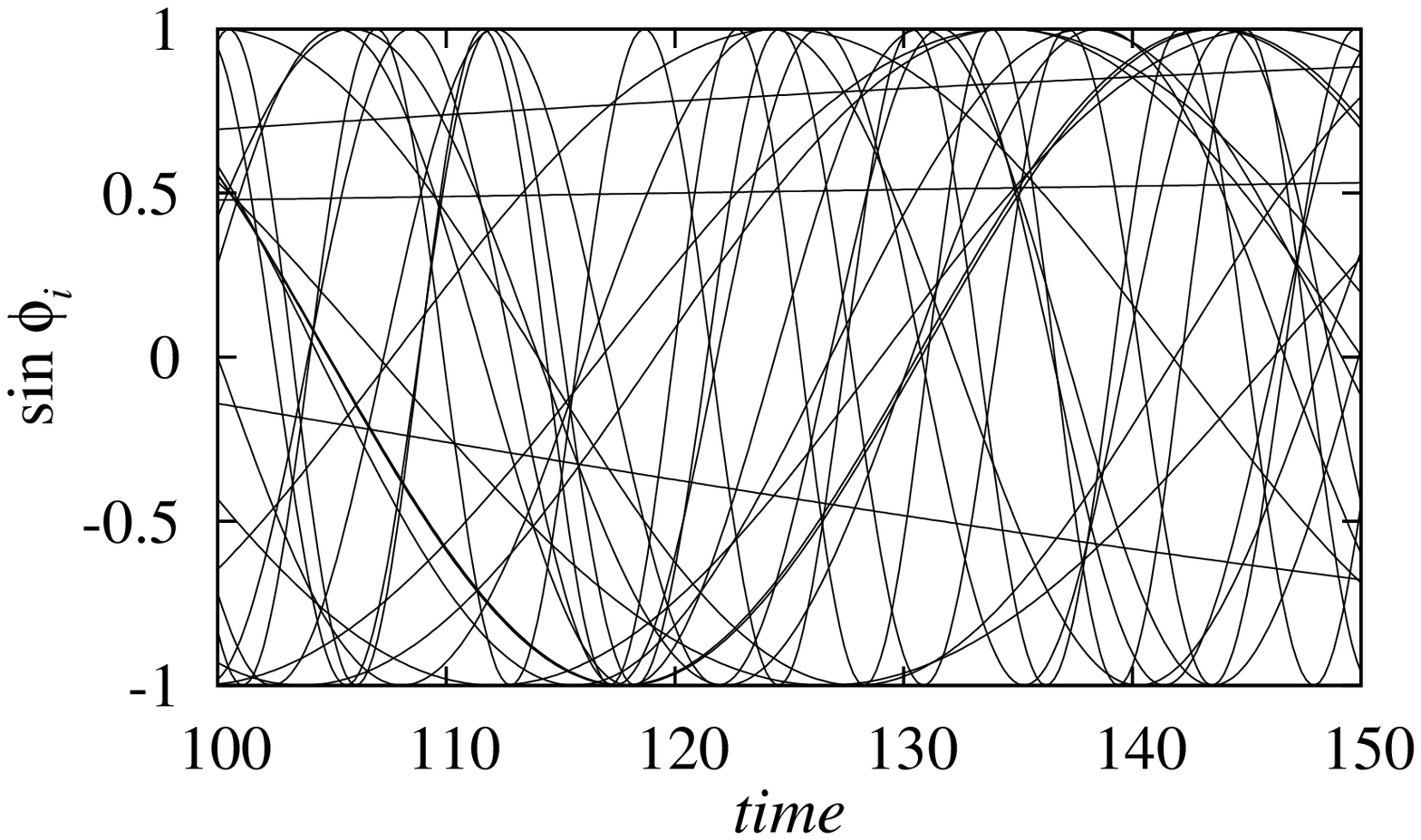}\\
     (b)\\
     \includegraphics[width=5.0cm]{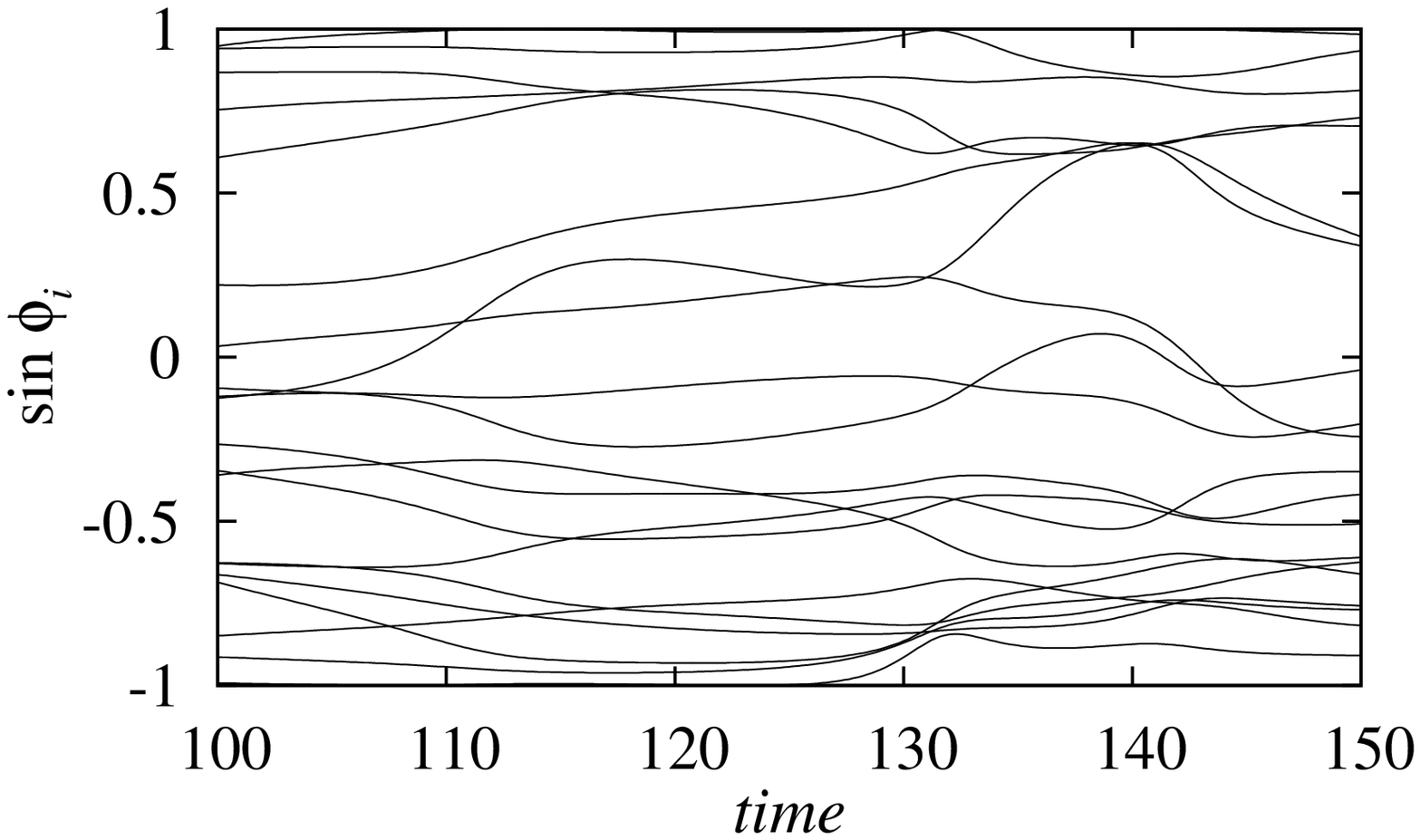}\\
     (c)\\
     \includegraphics[width=5.0cm]{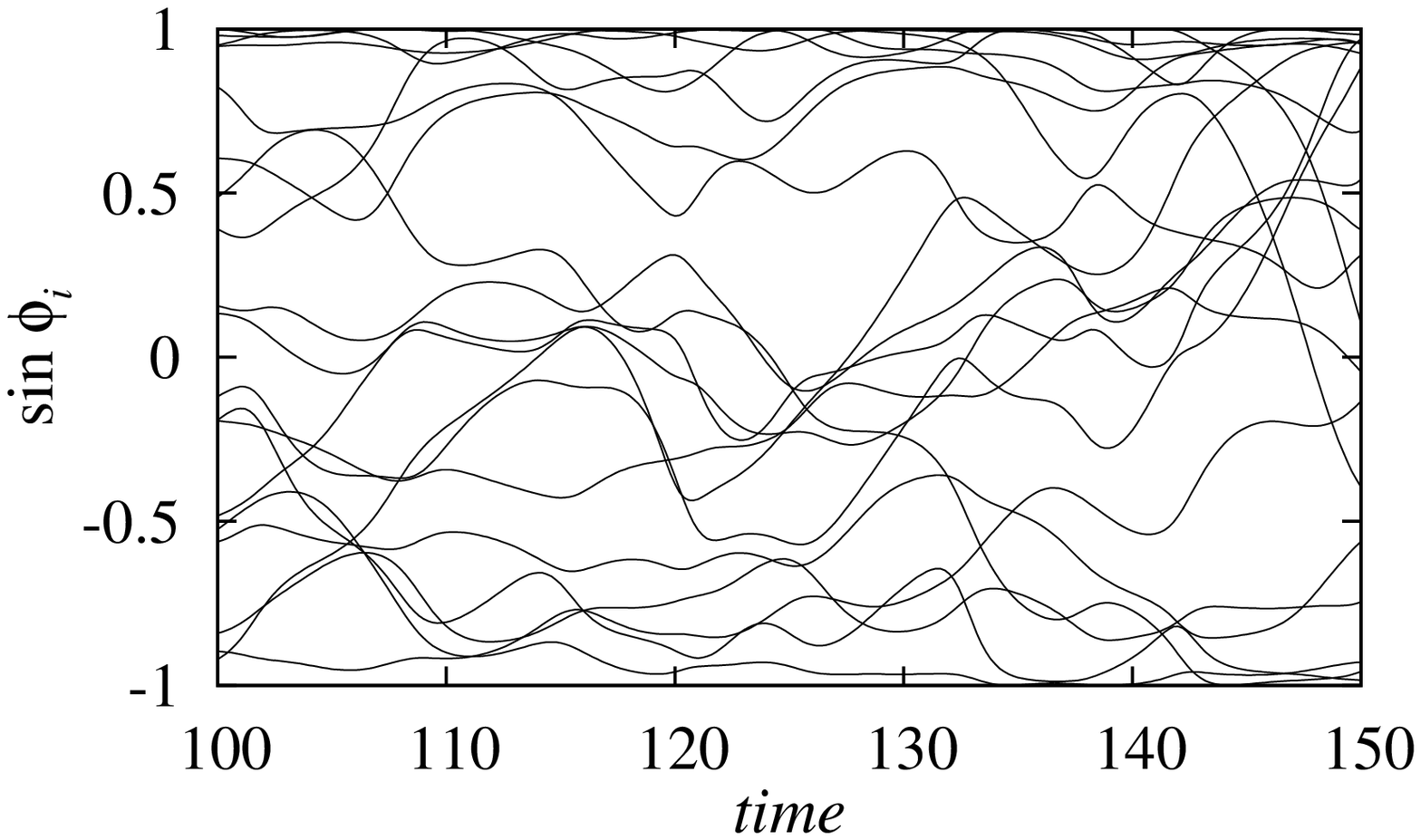}\\
     \if0
\includegraphics[width=5.0cm]{OSC(sigma=0.2)_test(Phi)_natural.eps}  \\
     \small(a) \\
     \includegraphics[width=5.0cm]{OSC(sigma=0.2)_test(Phi)_SK.eps} \\
   \small(b)  \\
     \includegraphics[width=5.0cm]{OSC(sigma=0.3)_test(Phi)_SK.eps} \\
     \small(c)
     \fi
    \end{tabular}
	\vspace{-3mm}
        \caption{Time series of $\sin \phi(t)$. $N=100$. 
           (a) $J_{jk}=0$, (b) $J_{jk} \ne 0, \sigma=0.2$, (c) $J_{jk} \ne 0, \sigma=0.3$.}
        %Temperature dependence of peak position of $P(r)$ in the XY model ($N=500$)
%  and $\sigma$ dependence of peak position of $P(r)$ in the phase oscillator network ($N=100$).}
  \label{fig:sinphi}
 \end{center}
\end{figure}

\subsection{Numerical results for several quantities in the phase oscillator network}
In the phase oscillator network,
in order to study the roles of synchronous and asynchronous oscillators 
for the correspondence, we numerically calculated  several quantities.
Firstly, we study the time evolution of $\sin \phi$ where $\phi$ is the
phase of each oscillator. %the instantaneous order parameter $q$.
In Fig. \ref{fig:sinphi}, we show %the time evolution of
 $\sin \phi(t)$ of 20 oscillators for $N=100$ during
$t=0 \sim 150$. In Fig. \ref{fig:sinphi}(a), we set $J_{jk}=0$, that is,
 $\phi_j=\omega_j t +\phi_j(0)$.  In Figs. \ref{fig:sinphi}(b) and (c),  
we set $J_{jk} \ne 0$, and $\sigma=0.2 \sqrt{\frac{\pi}{2}}$ and $0.3 \sqrt{\frac{\pi}{2}}$,
respectively.  We note that oscillators are locked for a while and then are unlocked, and
repeat this behavior. 
  We found that the larger $\sigma$ is, the more fluctuations of phases are,
and trajectories behave chaotically. Next,  we  studied trajectories of LFs for a
long time, from 0 to 2000 for $N=100 \sim  400$. See Figs. (\ref{fig:R}) and
 (\ref{fig:Theta}). %$[0, 2000]$. 
We define the amplitude $R_j$ and phase  $\Theta_j$ 
of the LFs by 
\bea
R_j e^{i \Theta_j}&=& p_j= \sum_k J_{jk} e^{i \phi_k}.
\eea
In this simulation, we adopted the simulated annealing
and the schedule is $T_l=0.7-(l-1)*0.02, l=1 \sim 35$.
 We obtained the following results.
 When $\sigma$ is small, $\sigma < \sigma_{c1}$,
 $R_j$  and   $\Theta_j$ are constant or periodic depending on $N$,
 where $\sigma_c \sim 0.1 \sqrt{\frac{\pi}{2}}$.
 The distribution of substantial frequencies is $G(\tilde{\omega})=\delta(\tilde{\omega})$.
  When $\sigma$ is large, $R_j$ behaves chaotically, and $\Theta_j$
 has two phases,
 in one phase
  $\Theta_j$ is almost constant, and in the other phase it increases or decreases drastically.
 On average, $\Theta_j$ evolves almost linearly.
 %The larger $N$ is, the shorter the
 % increases linearly in time
%for small $N$, but behaves intermittently for large $N$.
$G(\tilde{\omega})$ is one humped, continuous,  and it is impossible to separate synchronized
oscillators from desynchronized ones.
%It seems that  any oscillator has two phases of time,
%in one phase
%  $\phi_j$ is almost constant, and in the other phase it increases or decreases drastically.
%The larger $N$ is, the shorter the

In the next subsection, we derive the self-consistent equations for LFs
in the XY model and oscillator network by using approximations.
%when  $\sigma$ is small, that is, only the synchronized oscillators exist.
%For example, see Figs. \ref{fig:R} and  \ref{fig:Theta}
%\if0
%fig8 
\begin{figure}[H]
  %\begin{picture}(130,370)
\begin{picture}(100,250)  
\put(60,215){(a)}
%\put(5,115){\includegraphics[width=4.cm,clip]{R.lftime.n.100.ibeta.30.ismpl.1.place.1.20-100.t1800-2000.eps}}
\put(5,115){\includegraphics[width=4.cm,clip]{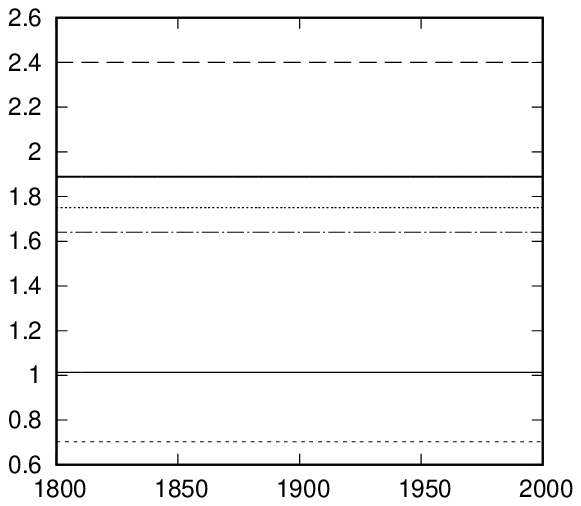}}
\put(70,105){$t$}
\put(-10,160){$R_j$}
\put(185,215){(b)}
%\put(130,115){\includegraphics[width=4.cm,clip]{R.lftime.n.100.ibeta.25.ismpl.1.place.1.20-100.t1800-2000.eps}}
\put(130,115){\includegraphics[width=4.cm,clip]{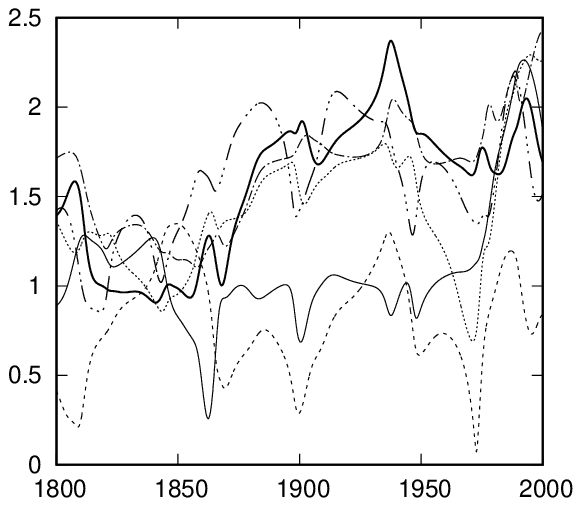}}
\put(185,105){$t$}
\put(120,160){$R_j$}
\put(60,90){(c)}
%\put(5,-10){\includegraphics[width=4.cm,clip]{R.lftime.n.100.ibeta.20.ismpl.1.place.1.20-100.t1800-2000.eps}}
\put(5,-10){\includegraphics[width=4.cm,clip]{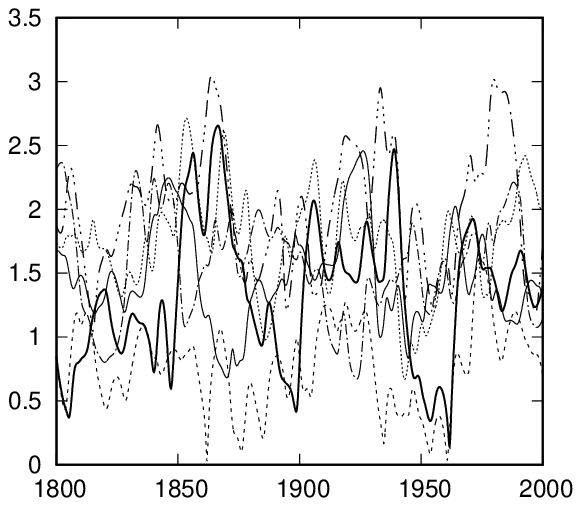}}
\put(70,-20){$t$}
\put(-10,35){$R_j$}
\put(185,90){(d)}
\put(130,-10){\includegraphics[width=4.cm,clip]{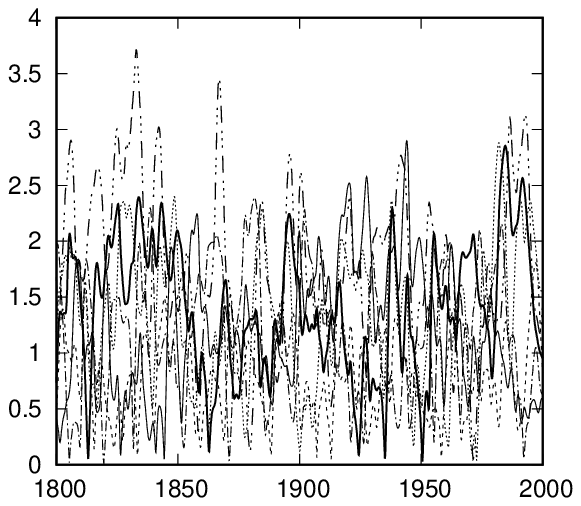}}
%\put(130,-10){\includegraphics[width=4.cm,clip]{R.lftime.n.100.ibeta.10.ismpl.1.place.1.20-100.t1800-2000.eps}}
\put(185,-20){$t$}
\put(120,35){$R_j$}
\end{picture}
\caption{Time series of  $R_j$s for several oscillators. $N=100$. 
  (a) $\sigma=0.1 \sqrt{0.5 \pi}$, 
(b) $\sigma=0.2 \sqrt{0.5 \pi}$,
(c) $\sigma=0.3 \sqrt{0.5 \pi}$,
  (d) $\sigma=0.5 \sqrt{0.5 \pi}$. 
}
\label{fig:R}
\end{figure}
%fig9 
\begin{figure}[H]
  %\begin{picture}(130,370)
\begin{picture}(100,250)  
\put(60,215){(a)}
\put(5,115){\includegraphics[width=4.cm,clip]{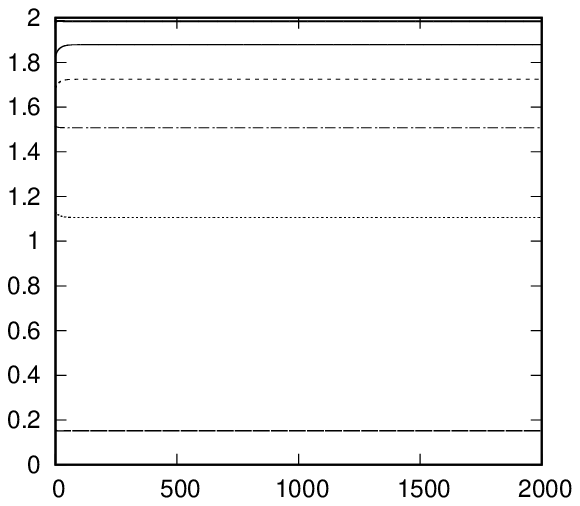}}
%\put(5,115){\includegraphics[width=4.cm,clip]{Theta.lftime.n.100.ibeta.30.ismpl.1.place.1.20-100.eps}}
\put(70,105){$t$}
\put(-10,160){$\Theta_j$}
\put(185,215){(b)}
\put(130,115){\includegraphics[width=4.cm,clip]{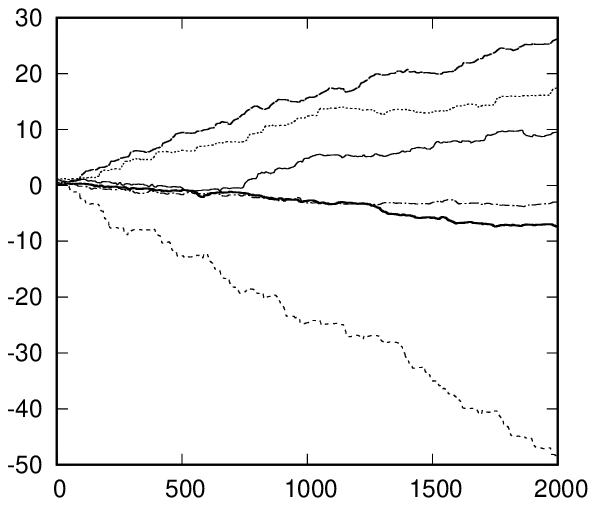}}
%\put(130,115){\includegraphics[width=4.cm,clip]{Theta.lftime.n.100.ibeta.25.ismpl.1.place.1.20-100.eps}}
\put(185,105){$t$}
\put(120,160){$\Theta_j$}
\put(60,90){(c)}
\put(5,-10){\includegraphics[width=4.cm,clip]{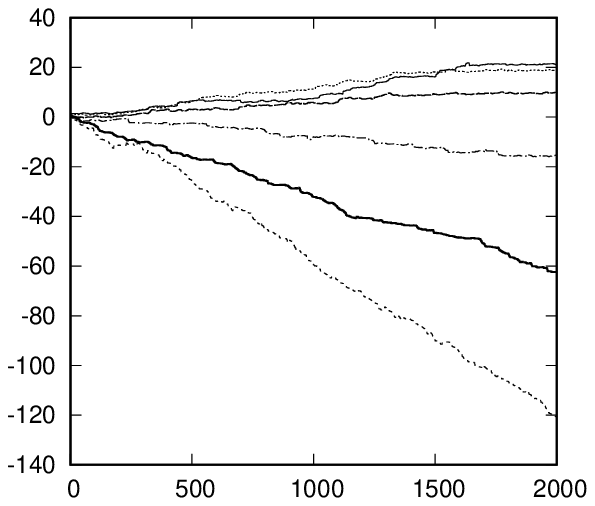}}
%\put(5,-10){\includegraphics[width=4.cm,clip]{Theta.lftime.n.100.ibeta.20.ismpl.1.place.1.20-100.eps}}
\put(70,-20){$t$}
\put(-10,35){$\Theta_j$}
\put(185,90){(d)}
\put(130,-10){\includegraphics[width=4.cm,clip]{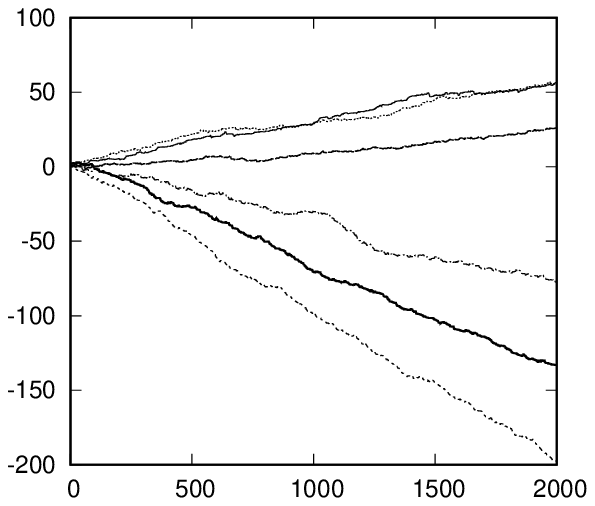}}
%\put(130,-10){\includegraphics[width=4.cm,clip]{Theta.lftime.n.100.ibeta.10.ismpl.1.place.1.20-100.eps}}
\put(185,-20){$t$}
\put(120,35){$\Theta_j$}
\end{picture}
\caption{Time series of  $\Theta_j$s for several oscillators. $N=100$.
  (a) $\sigma=0.1 \sqrt{0.5 \pi}$, 
(b) $\sigma=0.2 \sqrt{0.5 \pi}$,
(c) $\sigma=0.3 \sqrt{0.5 \pi}$,
  (d) $\sigma=0.5 \sqrt{0.5 \pi}$. 
}
\label{fig:Theta}
\end{figure}

\subsection{SCE for LFs}
In the oscillator network, we derive the SCE for the case that all oscillators are synchronized.
 In the XY model, we derive the SCE by using the naive mean-field approximation.
\subsubsection{Oscillator network}
Using  $R_j$ and $\Theta_j$, the evolution equation is rewritten as
\bea
\frac{d}{dt} \phi_j &=&\omega_j-R_j \sin (\phi_j - \Theta_j).
\eea
$R_j$ and $\Theta_j$ are constant because all oscillators
 are assumed to be synchronized. 
  Thus, by defining $\psi_j=\phi_j -\Theta_j$, we obtain
\bea
\frac{d}{dt} \psi_j &=&\omega_j-R_j \sin \psi_j.
\eea
The stable solution is $\psi_j ^*= \sin ^{-1}\frac{\omega_j}{R_j}$ where $|\psi_j ^*|<\frac{\pi}{2}$.
The probability density function of phases $f(\psi;\omega_j)$ is $\delta(\psi- \psi_j ^*)$.
Thus, the average of $e^{i \phi_j}$ is
\bea
\bra e^{i \phi_j}\ket &=& e^{i (\psi_j ^*+ \Theta_j)}
=\biggl(\sqrt{1-(\frac{\omega_j}{R_j})^2}+ i \frac{\omega_j}{R_j}\biggr)e^{i \Theta_j}.
\eea
Therefore, the SCE for LFs is
\bea
R_j e^{i \Theta_j} 
&=&\sum_{j' =1}^N J_{jj'}
\biggl(\sqrt{1-(\frac{\omega_{j'}}{R_{j'}})^2}+ i \frac{\omega_{j'}}{R_{j'}}\biggr)e^{i \Theta_{j'}}.
\label{I}
\eea
%We numerically solved the SCE (\ref{I}).
We numerically solved the SCE (\ref{I}) by iteration method.
That is, from  $\{ R_j \}$ and $\{ \Theta_j \}$ at time $t$, we evaluate the right-hand
 side of eq. (\ref{I}) to obtain $\{ R_j \}$ and $\{ \Theta_j \}$ at time $t+\Delta t$.
We define the distance between two configurations $\{\phi_j \}$ and $\{\phi_j ' \}$ as
\beas
d(\{\phi_j \}, \{\phi_j '  \}) \equiv \sum_{j=1}^N |\phi_j - \phi_j '|.
\eeas
The convergence condition is $ d(\{\phi_j(t) \}, \{\phi_j (t+\Delta t)  \}) < \epsilon$ 
for successive two configurations $\{\phi_j(t) \}$ and $\{\phi_j (t+\Delta t) \}$
with $\epsilon= 0.01$.
It turned out that it
is very difficult to obtain solutions for eq. (\ref{I}) if  initial conditions
are  taken randomly.
Then, as an initial condition, we used the numerical results
obtained by the simulated annealing method, and found that almost all  numerical results
are solutions of the SCE when $\sigma$ is small.
For example, we found that when $N=100$ and $\sigma=0.02 \sqrt{\frac{\pi}{2}}$,  
all 19 configurations obtained by the simulated annealing  converge by only one iteration
and  $ d(\{\phi_j (0)\}, \{\phi_j (\Delta t) \}) \sim 3 \times 10^{-5}$,
that is, these configurations satisfy eq. (\ref{I}).
We regard   two configurations $\{\phi_j \}$ and $\{\phi_j ' \}$ to be different when 
 $ d(\{\phi_j \}, \{\phi_j ' \}) > \epsilon$. 
We found only two different configurations  among 19 configurations. 
When $N=100$ and $\sigma=0.1 \sqrt{\frac{\pi}{2}}$,
we found that
 16 configurations  converge by only one iteration among 19 configurations, 
 and all of them are regarded as the same.
 However, for larger values of $\sigma$, we could not find any solution.
 This is because $R_j$ and $\Theta_j$ are not constant and it seems
  that asynchronous solutions contribute to the LFs.
 \subsubsection{XY model}
Hamiltonian is 
\bea
H&=& -\sum_{j<k} J_{jk}\cos(\phi_k-\phi_j)=%-\frac{1}{2} {\rm Re }\sum_{\j \ne k}
%J_{jk} e^{i (\phi_k-\phi_j)}\nonumber \\
%&=&-\frac{1}{2} {\rm Re }\sum_j e^{- i \phi_j}\biggl( \sum_ k J_{jk} e^{i \phi_k}  \biggr)
%=-\frac{1}{2} {\rm Re }\sum_j e^{-i \phi_j} R_j e^{i \Theta_j}\nonumber \\
%&=&
-\frac{1}{2} \sum_j R_j  \cos (\phi_j -\Theta_j). 
\eea
Since the probability density function of phases is
$P(\phi_j)=\frac{e^{\beta R_j \cos (\phi_j -\Theta_j) }}{2 \pi I_0(\beta R_j)}$,
 defining $\psi_j=\phi_j-\Theta_j$ we obtain
\bea
\bra e^{i \phi_j} \ket
&=&\frac{1}{2 \pi I_0(\beta R_j)}\int_0 ^{2 \pi}   e^{\beta R_j \cos \psi} e^{i (\psi+\Theta_j)}
d \phi \nonumber \\
&=&\frac{ I_1(\beta R_j)}{  I_0(\beta R_j)} e^{i \Theta_j}
= \beta R_j e^{i \Theta_j} u(\beta R_j).
 \eea
Here, $u(x)=\frac{I_1(x)}{x I_0(x)}$.
Thus, we obtain
\bea
R_j e^{i \Theta_j} &=& \sum_k J_{jk} \bra e^{i \phi_k} \ket
=\beta  \sum_k J_{jk} R_k e^{i \Theta_k} u(\beta R_k).
\label{eq.of.lf}
  \eea
  \if0
  したがって，
  \bea
  R_{jc}&=&R_j \cos \Theta_j =\beta  \sum_k J_{jk} R_{kc} u(\beta R_k),\\
  R_{js}&=&R_j \sin \Theta_j =\beta  \sum_k J_{jk} R_{ks} u(\beta R_k).
  \label{eqxy}
  \eea
  \fi
  As an initial condition, we used the configuration obtained by the simulated annealing
  as in the oscillator network.  The method to solve eq. (\ref{eq.of.lf}),
  the convergence condition, and   the criterion of different solutions are the
  same as in the phase oscillator network. 
  When $N=100$ and $T=0.02$, among 30 configurations, 3 configurations converge with
  $\epsilon=0.01$. The numbers of iterations are rather large compared to 
   the oscillator network, and are 
  29, 51, and 62 for these three configurations, respectively.
   All of them are different, but it is difficult to
  distinguish these three from the figure of $j$ vs. $R_j$.
  When $N=100$ and $T=0.1$, among 30 configurations, 5 configurations converge, 
  %  and $ d(\{\phi_j \}, \{\phi_j ' \}) $ are order of 0.01.
   and the number of iterations ranges from 50 to 70.
   Four configurations among 5 are different.
  We found that convergent values and initial conditions are rather different
  and this is consistent with the fact that  the numbers of iterations are large.
 See Fig. \ref{jvsRj}. 
  Therefore, in this case, final configurations by the simulated annealing for $T=0.1$ are
  not considered as the solutions of the SCEs. 
  The reason for this is considered that the naive mean field approximation is not valid
   for the high temperatures.
   %See Fig. \ref{jvsRj}).
%fig10
 \begin{figure}[H]
\begin{picture}(130,370)
\put(110,340){(a)}                           
\put(0,180){\includegraphics[width=8.cm,clip]{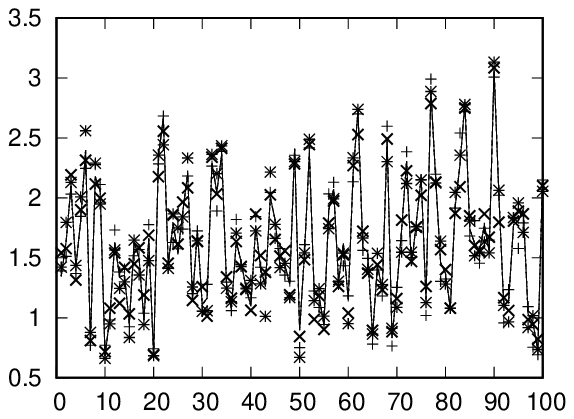}}
%\put(0,180){\includegraphics[width=8.cm,clip]{ib35.xy.sim.difsol.theory.rr.n.100.ibeta.34.ismpl.1.4.23.eps}}
\put(120,175){$j$}
\put(-8,260){$R_j$}
\put(1100,165){(b)} 
\put(0,-5){\includegraphics[width=8.cm,clip]{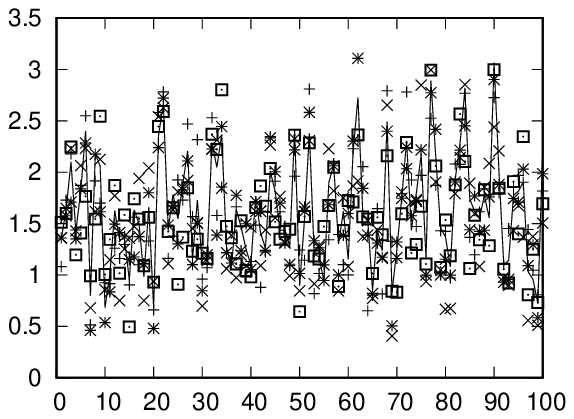}}
%\put(0,-5){\includegraphics[width=8.cm,clip]{ib35.xy.sim.difsol.theory.rr.n.100.ibeta.30.ismpl.1.10.17.29.eps}}
\put(120,-10){$j$}
\put(-8,75){$R_j$}
\end{picture}
\caption{$j$ dependences of $R_j$. XY model. $N=100$.
  Symbols: different solutions among convergent solutions
  obtained by the iteration of eq. (\ref{eq.of.lf}), broken line:
  initial condition which is the final value of the annealing.
  (a) $T=0.02$, 3 different solutions.   (b) $T=0.1$, 4 different solutions.
}
\label{jvsRj}
\end{figure}
 \if0
\begin{figure}[H]
 \begin{center}
   \begin{tabular}{c}
	\small(a) \\
      \includegraphics[width=8.cm,clip]{ib35.xy.sim.difsol.theory.rr.n.100.ibeta.34.ismpl.1.4.23.eps}\\
	\small(b) \\
\includegraphics[width=8.cm,clip]{ib35.xy.sim.difsol.theory.rr.n.100.ibeta.30.ismpl.1.10.17.29.eps}\\
    \end{tabular}
\caption{$j$ dependences of $R_j$. XY model. $N=100$.
  Symbols: different solutions among convergent solutions
  obtained by the iteration of eq. (\ref{eq.of.lf}), broken line:
  initial condition which is the final value of the annealing.
  (a) $T=0.02$, 3 different solutions.   (b) $T=0.1$, 4 different solutions.
}
\label{jvsRj}
 \end{center}
\end{figure}
\fi
% \end{document}
  \section{Summary and discussion}
  We summarize the results of this paper.
  We studied the random and frustrated interaction, the SK interaction, which
   is generated by the Gaussian distribution with mean 0 and standard deviation $J/\sqrt{N}$.
   As for the distribution of natural frequencies $g(\omega)$,
   we adopted the Gaussian distribution with mean 0 and standard deviation  $\sigma$.
  In order to study whether correspondence between the two models 
  exists or not,  we performed numerical calculations
of the spin glass order parameter $q$ and the distributions of local fields (LFs)
 in the XY model and phase oscillator network. 
 In the XY model, we used the Markov Chain Monte Carlo simulation (MCMCs),
  in particular, the replica exchange Monte Carlo (REMC) method and 
  the simulated annealing method. In the oscillator network, we used
  the Euler method with time increment $\Delta t = 0.02$, and also used the
  simulated  annealing method, that is, we integrate the evolution equation by decreasing
  $\sigma$ slowly.
  First, we summarize the results of $q$.
  In the XY model, we confirmed that theoretical and numerical
  results agree fairly well  and found that
  the coinciding region between  the  theoretical curve 
  and the simulation results of $q$ increases as $N$ increases. 
 For the phase oscillator network, we  found that in the $\sigma$  dependence of
  the spin glass order parameter 
  the coinciding region between  the  theoretical curve $q(T(\sigma))$
of the XY model and the simulation results
decreases as $N$ increases, contrary to our expectation.
  Here,  $T(\sigma)=\sqrt{\frac{2}{\pi}}\sigma$
  is the relation obtained in the previous paper.
  Since $\phi_j$ behaves intermittently in time, we introduced the 
  order parameter $q_{\rm av}$ for the time averaged phases, and found that
  the coinciding region between  the  theoretical curve 
of the XY model and the simulation results of $q_{\rm av}$ 
increases as $N$ increases.\\
 Next, we summarize the results of LFs.
% The distributions of local fields of the two models behave similarly.\\
% Thus, the quasientrainment state in the oscillator network
 % appears to correspond to the spin glass state in the XY model.
 We define the probability density $P(r)$ of LFs, where $r$ is the
 radius of the local field in the complex plane.
 As $T$ or $\sigma$ increases, the peak radius $r_p$ of $P(r)$ changes from non-zero
 value to 0. This is the so called volcano transition, and the
 transition points of the two models seem to  correspond according to the relation
  $T=\sqrt{\frac{2}{\pi}} \sigma$.
 For the oscillator network, we numerically studied time evolution of $\sin \phi_i$
  of each oscillator 
 and found that oscillators are locked for a while and then are unlocked,
  and repeat this behavior.  We also numerically studied time evolution amplitudes $R$s and 
  phases $\Theta$s of LFs. 
 We found that   when $\sigma$ is small, 
  they are constant or periodic
  depending on $N$, and the distribution of the substantial frequencies $G(\omega)$ is the delta
  function $\delta(\omega)$, 
   but when $\sigma$ is large, $R_j$ behaves chaotically,
%  The phases increase linearly on average in time for small $N$ but behaves intermittently for large
   and $\Theta_j$ has two phases, in one phase
  $\Theta_j$ is almost constant, and in the other phase it increases or decreases drastically.
   On average, $\Theta_j$ evolves almost linearly.
    $G(\omega)$ is one-humped and continuous. \\
  Finally, we derived the self-consistent equation (SCE) of LFs for 
  the oscillator network in the case that all oscillators synchronize, and
  for the  XY model  by using the naive mean field approximation.
  We found that for the oscillator network  and XY model, when  $\sigma$ and $T$ are small,
  configurations  obtained by simulated annealing satisfy the SCE, 
  but when $\sigma$ and $T$ are large, they do not.
    The reasons for the discrepancy between theoretical and numerical
  results for the LFs at large $T$ and $\sigma$  are considered as follows.
  In the  oscillator network, the asynchronous oscillators do not  contribute  to the LFs
  for the solvable models when the $g(\omega)$ is one-humped and symmetric with respect to
   its center.  However, the present results imply that asynchronous oscillators contribute
  to the LFs. Since $G(\omega)$ is continuous, it is difficult to separate
  synchronized oscillators from desynchronized ones. 
  In the XY model, the present results imply that the naive mean-field approximation is
  not valid except for very low temperatures.
  This is the same as in the case of Ising spins.
   The so called Onsager reaction field should be
   taken into account for the XY model as in the Ising model.
   Therefore, in order to improve the present approximations for the two models,
    further elaborate studies are necessary, and these studies 
    are  beyond the scope of the present paper    and are left as  a future problem.\\
    
% These results are considered to be strong evidence
% that correspondence exists between the two models.\par
%  A future problem related to the  present study
% is to directly and theoretically prove  correspondence between the two models
 %  in the SK interaction.
 \if0
More general future problems are
 to clarify whether  correspondence exists or not between the two models 
 depending on  dimensionality, interaction range, types of randomness, topology of
 the system, and so on. Furthermore, it is an interesting problem to
 study  correspondence of critical phenomena.\par
 \fi
  The present study is
   supported by JPSJ KAKENHI Grant No. 16K05474, No. 25330298,
   No. 17K00357.

\section{Appendix A}
In this appendix, we derive the disorder averaged free energy per spin and the SPEs under the ansatz of the replica symmetry.
 The derivation  is based on Ref. \cite{Hatchett.Uezu.2008}
\bea
\bar{f}&=&-\lim _{N \to \infty}\lim _{n \to 0}
(\beta N n )^{-1} \log \int d \bphi^1 \cdots \bphi^n
\overline{e^{-\beta \sum_{\alpha} H(\bphi^\alpha)}},\\
\overline{e^{-\beta \sum_{\alpha} H(\bphi^\alpha)}}&=&
\exp[\frac{\beta ^2 J^2}{4N} \sum_{\alpha \beta}
  \{ (\sum_i \cos \phi_i ^\alpha \cos \phi_i ^\beta)^2
  +   (\sum_i \cos \phi_i ^\alpha \sin \phi_i ^\beta)^2  \nonumber \\
&&+   (\sum_i \cos \phi_i ^\alpha \sin \phi_i ^\beta)^2 +
       (\sum_i \sin \phi_i ^\alpha \cos \phi_i ^\beta)^2 +
  -N \}],
\eea
where $\bphi^\alpha = (\phi_1 ^\alpha, \cdots, \phi_N ^\alpha)$.
We define the following order parameters.
For $\alpha< \beta$, 
\bea
&& q_{\rmcc}^{\alpha \beta}=\frac{1}{N}\sum_i \cos \phi_i ^\alpha \cos \phi_i ^\beta, \
q_{\rmss}^{\alpha \beta}=\frac{1}{N}\sum_i \sin \phi_i ^\alpha \sin \phi_i ^\beta, \nonumber \\
&&q_{\rmcs}^{\alpha \beta}=\frac{1}{N}\sum_i \cos \phi_i ^\alpha \sin \phi_i ^\beta, \
q_{\rmsc}^{\alpha \beta}=\frac{1}{N}\sum_i \sin \phi_i ^\alpha \cos \phi_i ^\beta, \nonumber
\eea
and for $\alpha=1, \cdots,n$, 
\bea
&& Q_{\rmcc}^{\alpha}=\frac{1}{N}\sum_i \cos^2 \phi_i ^\alpha, \ 
Q_{\rmss}^{\alpha}=\frac{1}{N}\sum_i \sin ^2 \phi_i ^\alpha, \
Q_{\rmcs}^{\alpha}=\frac{1}{N}\sum_i \cos \phi_i ^\alpha \sin \phi_i ^\alpha.\nonumber 
\eea
Then, we obtain
\bea
\overline{e^{-\beta \sum_{\alpha} H(\bphi^\alpha)}}&=&
e^{-\frac{\beta ^2 J^2 n^2}{4}} \exp[\frac{\beta ^2 J^2N}{4} \sum_{\alpha}
  \{(Q_{\rmcc} ^\alpha )^2+(Q_{\rmss} ^\alpha )^2+2(Q_{\rmcs} ^\alpha )^2 \} \nonumber \\
&&  + 2  \sum_{\alpha <\beta} \{
  (q_{\rmcc} ^{\alpha  \beta})^2+  (q_{\rmss} ^{\alpha  \beta})^2 
       +   (q_{\rmcs} ^{\alpha  \beta})^2 + (q_{\rmsc} ^{\alpha  \beta})^2 \}].
\eea
Using the integral representation of $\delta$ functions such as
\beas
\int \frac{1}{2 \pi} dq_{\rmcc} ^{\alpha \beta} d \hatq_{\rmcc} ^{\alpha \beta}
e^{ i \hatq_{\rmcc} ^{\alpha \beta}(q_{\rmcc} ^{\alpha \beta} -\frac{1}{N} \sum_i \cos \phi_i ^\alpha
  \cos \phi_i ^\beta)}=1,
\eeas
and re-scaling variables as $\hatq_{\rmcc}^{\alpha \beta} \to N \hatq_{\rmcc}^{\alpha \beta}$,
$\hatQ_{\rmcc}^{\alpha} \to N \hatQ_{\rmcc}^{\alpha}$, etc., 
we obtain
\bea
\bar{f}&=&-\lim _{N \to \infty}\lim _{n \to 0}
(\beta N n )^{-1} \log \{ \int d \bq e^{N(\Phi+\Psi)} \},\\
\Phi&=& 
  i \sum_{\alpha} \{ \hatQ_{\rmcc}^{\alpha} Q_{\rmcc}^{\alpha}
  +\hatQ_{\rmss}^{\alpha} Q_{\rmss}^{\alpha}
  +\hatQ_{\rmcs}^{\alpha} Q_{\rmcs}^{\alpha}
  \}
  + i \sum_{\alpha<\beta}
  \{ \hatq_{\rmcc}^{\alpha \beta} q_{\rmcc}^{\alpha \beta}
  +\hatq_{\rmss}^{\alpha \beta} q_{\rmss}^{\alpha \beta}
  +   \hatq_{\rmcs}^{\alpha \beta} + \hatq_{\rmsc}^{\alpha \beta}
  \} \nonumber \\
&&  +\frac{\beta ^2 J^2}{4}\biggl\{
 \sum_{\alpha} \biggl( (Q_{\rmcc}^{\alpha})^2
 +(Q_{\rmss}^{\alpha})^2  +2(Q_{\rmcs}^{\alpha})^2 \biggr)
+ 2 \sum_{\alpha<\beta} \biggl((q_{\rmcc}^{\alpha \beta})^2  +(q_{\rmss}^{\alpha \beta})^2 
+  ( q_{\rmcs}^{\alpha \beta})^2+  ( q_{\rmsc}^{\alpha \beta})^2 \biggr)
\biggr\},\nonumber \\
  \Psi&=& \frac{1}{N}\log \biggl\{
      [\int \prod_\alpha d \bphi ^\alpha]\nonumber \\
      &&   \times   \exp [
        -i \sum_{\alpha} \sum_i
        ( \hatQ_{\rmcc} ^{\alpha \beta} \cos ^2\phi_i ^\alpha
        +\hatQ_{\rmss} ^{\alpha \beta} \sin ^2 \phi_i ^\alpha
        +\hatQ_{\rmcs} ^{\alpha \beta} \cos \phi_i ^\alpha \sin  \phi_i ^\alpha)         \nonumber \\
&&        -i \sum_{\alpha<\beta} \sum_i
        ( \hatq_{\rmcc} ^{\alpha \beta} \cos \phi_i ^\alpha \cos \phi_i ^\beta
        + \hatq_{\rmss} ^{\alpha \beta} \sin \phi_i ^\alpha \sin \phi_i ^\beta
        + \hatq_{\rmcs} ^{\alpha \beta} \cos \phi_i ^\alpha \sin \phi_i ^\beta
      + \hatq_{\rmsc} ^{\alpha \beta} \sin \phi_i ^\alpha \cos \phi_i ^\beta        
      ]
      \biggr\},\nonumber \\
      \eea
      where $ d \bq =
      \prod_{\alpha} \biggl(
      \frac{N d \hatQ_{\rmcc}^{\alpha } d Q_{\rmcc}^{\alpha }}{2 \pi}
      \frac{N d \hatQ_{\rmss}^{\alpha } d Q_{\rmss}^{\alpha}}{2 \pi}
      \frac{N d \hatQ_{\rmcs}^{\alpha } d Q_{\rmcs}^{\alpha }}{2 \pi}
        \biggr)
      \prod_{\alpha<\beta} \biggl(
      \frac{N d \hatq_{\rmcc}^{\alpha \beta} d q_{\rmcc}^{\alpha \beta}}{2 \pi}
      \frac{N d \hatq_{\rmss}^{\alpha \beta} d q_{\rmss}^{\alpha \beta}}{2 \pi}     
      \frac{N d \hatq_{\rmcs}^{\alpha \beta} d q_{\rmcs}^{\alpha \beta}}{2 \pi}
      \frac{N d \hatq_{\rmsc}^{\alpha \beta} d q_{\rmcs}^{\alpha \beta}}{2 \pi}      
       \biggr)$.
      Since we consider $N \to \infty$, the integration is estimated by the saddle point
      of $\Phi + \Psi$,
      \bea
\bar{f}&=&-\lim _{N \to \infty}\lim _{n \to 0}
(\beta N n )^{-1} \log \{ \int d \bq e^{N(\Phi+\Psi)} \}
\sim -\lim _{n \to 0} (\beta  n )^{-1} \mbox{ extr }(\Phi+\Psi).
\eea
Now, let us consider the replica symmetric solution.
\bea
&& Q_{\rmcc}^{\alpha} =Q_{\rmcc}, \ Q_{\rmss}^{\alpha} =Q_{\rmss}, \
Q_{\rmcs}^{\alpha} =Q_{\rmcs}, \
 \hatQ_{\rmcc}^{\alpha} =\hatQ_{\rmcc}, \ \hatQ_{\rmss}^{\alpha} =\hatQ_{\rmss}, \
\hatQ_{\rmcs}^{\alpha} =\hatQ_{\rmcs}, \\
&& q_{\rmcc}^{\alpha \beta} =q_{\rmcc}, \ q_{\rmss}^{\alpha \beta} =q_{\rmss}, \
q_{\rmcs}^{\alpha \beta} =q_{\rmcs}, \ 
 \hatq_{\rmcc}^{\alpha \beta} =\hatq_{\rmcc}, \ \hatq_{\rmss}^{\alpha \beta} =\hatq_{\rmss}, \
 \hatq_{\rmcs}^{\alpha \beta} =\hatq_{\rmcs}, \hatq_{\rmsc}^{\alpha \beta} =\hatq_{\rmsc}.\nonumber \\
\eea
Then , by changing conjugate variables from $\hatq_{\rmcc} \to i \hatq_{\rmcc},
\hatQ_{\rmcc} \to i \hatQ_{\rmcc}$, etc.,  we obtain
\bea
\lim _{n \to 0} \frac{1}{n}
\Phi_{\rm RS} &=&
  - ( \hatQ_{\rmcc} Q_{\rmcc}  +\hatQ_{\rmss} Q_{\rmss}
  +\hatQ_{\rmcs} Q_{\rmcs})
  +\frac{1}{2} ( \hatq_{\rmcc} q_{\rmcc}
  +\hatq_{\rmss} q_{\rmss}  + \hatq_{\rmcs} q_{\rmcs}+ \hatq_{\rmsc} q_{\rmsc}
  ) \nonumber \\
&&  +\frac{\beta ^2 J^2}{4}( Q_{\rmcc}^2
 +Q_{\rmss}^2  +2 Q_{\rmcs}^2 
 -q_{\rmcc}^2 -q_{\rmss}^2  - q_{\rmcs}^2- q_{\rmsc}^2),\\
 \lim _{n \to 0}  \frac{1}{n} \Psi_{\rm RS} & = &
 \lim _{n \to 0} \frac{1}{n} \log ( \int[ \prod_\alpha d \phi^\alpha]
 e^L ),\\
L&=& \sum_{\alpha} ( \hatQ_{\rmcc} \cos ^2\phi ^\alpha
        +\hatQ_{\rmss}  \sin ^2 \phi ^\alpha
        +\hatQ_{\rmcs} \cos \phi ^\alpha \sin  \phi ^\alpha)        \nonumber \\
&&        + \sum_{\alpha<\beta} 
        ( \hatq_{\rmcc}  \cos \phi ^\alpha \cos \phi ^\beta
        + \hatq_{\rmss}  \sin \phi ^\alpha \sin \phi ^\beta
        + \hatq_{\rmcs}  \cos \phi ^\alpha \sin \phi ^\beta
        + \hatq_{\rmsc}  \sin \phi ^\alpha \cos \phi ^\beta).        \nonumber \\
        \eea
        By using the Hubbard-Stratonovich transformation, $e^L$ is rewritten as
        \bea
        e^L&=&
   \exp[\biggl(\hatQ_{\rmcc} -\frac{1}{2} \hatq_{\rmcc}\biggr) \sum_\alpha \cos ^2 \phi^\alpha
     +\biggl(\hatQ_{\rmss} -\frac{1}{2} \hatq_{\rmss}\biggr) \sum_\alpha \sin^2 \phi^\alpha\nonumber \\
     &&     +\biggl(\hatQ_{\rmcs} - \frac{\hatq_{\rmcs}+\hatq_{\rmsc}}{2}   \biggr)
     \sum_\alpha \sin \phi^\alpha \cos \phi^\alpha]\nonumber \\
   && \times \int Dx \int Dy \nonumber \\
   && \times \exp[
     \sqrt{\frac{\hatq_{\rmc}\hatq_{\rmss}-(\frac{\hatq_{\rmcs}+\hatq_{\rmcs}}{2})^2}{\hatq_{\rmss}}}     
       \sum_{\alpha}\cos \phi^\alpha \ x
       +\biggl(\frac{(\hatq_{\rmcs}+\hatq_{\rmsc})}{2\sqrt{\hatq_{\rmss}}}
       \sum_\alpha \cos \phi^\alpha
       +\sqrt{\hatq_{\rmss}} \sum_{\alpha}\sin \phi^\alpha \biggr)y], \nonumber \\
       \eea
   Then, we obtain
   \bea
   \lim _{n \to 0}  \frac{1}{n} \Psi_{\rm RS} & = &
   \int Dx \int Dy \log \int d \phi
   \exp \biggl[\biggl(\hatQ_{\rmcc} -\frac{1}{2}\hatq_{\rmcc}  \biggr)  \cos ^2 \phi
     +\biggl(\hatQ_{\rmss} -\frac{1}{2} \hatq_{\rmss}\biggr) \sin^2 \phi  \nonumber \\
     &&     +\biggl(\hatQ_{\rmcs} - \frac{\hatq_{\rmcs}+\hatq_{\rmsc}}{2}  \biggr)
     \sin \phi \cos \phi \nonumber \\
&& +
       \sqrt{\frac{\hatq_{\rmc}\hatq_{\rmss}-(\frac{\hatq_{\rmcs}+\hatq_{\rmcs}}{2})^2}{\hatq_{\rmss}}}
       \cos \phi \ x
       +\biggl(\frac{\hatq_{\rmcs}+\hatq_{\rmsc}}{2\sqrt{\hatq_{\rmss}}}\cos \phi
       +\sqrt{\hatq_{\rmss}} \sum_{\alpha}\sin \phi \biggr) y\biggr]. \nonumber \\
\eea
   $\bar{f}_{\rm RS}$ is expressed as
\bea
\bar{f}_{\rm RS} &=& -\frac{1}{\beta}
\lim _{n \to 0}  \frac{1}{n}( \Phi_{\rm RS} +\Psi_{\rm RS} ) \nonumber \\
&=& -\frac{1}{\beta}\biggl\{ - ( \hatQ_{\rmcc} Q_{\rmcc}  +\hatQ_{\rmss} Q_{\rmss}
  +\hatQ_{\rmcs} Q_{\rmcs})
  +\frac{1}{2} ( \hatq_{\rmcc} q_{\rmcc}
  +\hatq_{\rmss} q_{\rmss}  + \hatq_{\rmcs} q_{\rmcs}+ \hatq_{\rmsc} q_{\rmsc}  ) \nonumber \\
&&  +\frac{\beta ^2 J^2}{4}( Q_{\rmcc}^2
 +Q_{\rmss}^2  +2 Q_{\rmcs}^2 
 -q_{\rmcc}^2 -q_{\rmss}^2-  q_{\rmcs}^2 -  q_{\rmss}^2
 ),\nonumber \\
 &&+  \int Dx \int Dy \log \int d \phi
   \exp \biggl[\biggl(\hatQ_{\rmcc} -\frac{1}{2} \hatq_{\rmcc}\biggr)  \cos ^2 \phi
     +\biggl(\hatQ_{\rmss} -\frac{1}{2} \hatq_{\rmss}\biggr) \sin^2 \phi  \nonumber \\
     &&     +\biggl(\hatQ_{\rmcs} - \frac{ \hatq_{\rmcs}+\hatq_{\rmcs}}{2}
       \biggr)
     \sin \phi \cos \phi \nonumber \\
&& +
     \sqrt{\frac{\hatq_{\rmcc}\hatq_{\rmss}-(\frac{\hatq_{\rmcs}+\hatq_{\rmsc}}{2} )^2}{\hatq_{\rmss}}}     
       \cos \phi \ x
       +\biggl(\frac{\hatq_{\rmcs}+\hatq_{\rmsc}}{2\sqrt{\hatq_{\rmss}}}\cos \phi
       +\sqrt{\hatq_{\rmss}} \sum_{\alpha}\sin \phi \biggr)y\biggr] \biggr\}. \nonumber \\
   \eea
   From the extrema conditions with respect to
   $q_{\rmcc}, q_{\rmss}, q_{\rmcs}, q_{\rmsc}$ and $ Q_{\rmcc}, Q_{\rmss}, Q_{\rmcs}$, we obtain
   \bea
   && \hatq_{\rmcc}=\beta ^2 J^2 q_{\rmcc}, \ \hatq_{\rmss}=\beta ^2 J^2 q_{\rmss}, \
   \hatq_{\rmcs}=\beta ^2 J^2 q_{\rmcs}, \ \hatq_{\rmsc}=\beta ^2 J^2 q_{\rmsc},
   \nonumber \\
   && \hatQ_{\rmcc}=\frac{\beta ^2 J^2}{2} Q_{\rmcc}, \
   \hatQ_{\rmss}=\frac{\beta ^2 J^2}{2} Q_{\rmss}, \
   \hatQ_{\rmcs}=\beta ^2 J^2 Q_{\rmcs}.
   \eea
   Thus, we have
   \bea
   \bar{f}_{\rm RS} &=& -\frac{1}{\beta}
   \biggl\{
   \frac{\beta ^2 J^2}{4} (
   q_{\rmcc}^2+q_{\rmss}^2+q_{\rmcs}^2+q_{\rmsc}^2
 -Q_{\rmcc}^2-Q_{\rmss}^2 -2Q_{\rmcs}^2)\nonumber \\
&&  +\int Dx \int Dy \log \int d \phi M(\phi | x,y) \biggr\}, \\
 M(\phi| x,y)&=& \exp\biggl[
 \frac{\beta ^2 J^2}{2} (  Q_{\rmcc}  - q_{\rmcc}) \cos ^2 \phi+
   \frac{\beta ^2 J^2}{2} (  Q_{\rmss}  - q_{\rmss}) \sin ^2 \phi \nonumber \\
   &&+ \beta ^2 J^2 (  Q_{\rmcs}  -
   \frac{q_{\rmcs}+q_{\rmsc}}{2})
    \sin \phi  \cos \phi \nonumber \\
&& + \beta J
    \sqrt{\frac{q_{\rmcc} q_{\rmss}-(
\frac{     q_{\rmcs}+     q_{\rmsc}}{2})^2}{q_{\rmss}}}     
       \cos \phi \ x
       + \beta J \biggl(\frac{q_{\rmcs}+q_{\rmsc}       }{2\sqrt{q_{\rmss}}}\cos \phi
       +\sqrt{q_{\rmss}} \sin \phi \biggr)y \biggr].\nonumber \\
 \eea
 From this, we obtain the following SPEs.
 \bea
 && Q_{\rmcc}=[\bra \cos ^2 \phi \ket], Q_{\rmss}=[\bra \sin ^2 \phi \ket]=1-Q_{\rmcc}, \
 Q_{\rmcs}=[\bra \sin \phi  \cos \phi \ket],\\
 && q_{\rmcc}=[\bra \cos \phi \ket^2], q_{\rmss}=[\bra \sin \phi \ket^2], \
 q_{\rmcs}=[\bra \sin \phi \ket \bra \cos \phi \ket]=q_{\rmsc},\\
 && [\cdots]\equiv \int Dx \int Dy \cdots, \ \bra \cdots \ket \equiv \frac{\int d \phi M(\phi|x,y)
   \ \cdots }{\int d \phi M(\phi|x,y)}.
 \eea
 Using above relations, $f_{\rm RS}$ and $L(\phi|x,y)$ are now expressed as
 \bea
 \bar{f}_{\rm RS} &=& -
   \frac{\beta  J^2}{4} \biggl(
   q_{\rmcc}^2+q_{\rmss}^2+2 q_{\rmcs}^2
 -1 +2Q_{\rmcc}(1-Q_{\rmcc}) -2Q_{\rmcs}^2 \biggr)\nonumber \\
&& -\frac{1}{\beta} \int Dx \int Dy \log \int d \phi M(\phi | x,y),\\
 M(\phi| x,y)&=& \exp\biggl[
 \frac{\beta ^2 J^2}{2} (  Q_{\rmcc}  - q_{\rmcc}) \cos ^2 \phi+
   \frac{\beta ^2 J^2}{2} ( 1- Q_{\rmcc}  - q_{\rmss}) \sin ^2 \phi \nonumber \\
&&+ \beta ^2 J^2 (  Q_{\rmcs}  - q_{\rmcs}) \sin \phi  \cos \phi \nonumber \\
&& + \beta J
\sqrt{\frac{q_{\rmcc} q_{\rmss}-q_{\rmcs}^2}{q_{\rmss}}}     
       \cos \phi \ x
       + \beta J \biggl(\frac{ q_{\rmcs}}{\sqrt{q_{\rmss}}}\cos \phi
       +\sqrt{q_{\rmss}} \sin \phi \biggr)y \biggr].
 \eea
 From the results by the simulated annealing, we assume
 $Q_{\rmcc}= Q_{\rmss}=\frac{1}{2}$, and $Q_{\rmcs}=0$.
 Then, 
 \bea
 \bar{f}_{\rm RS} &=& -
   \frac{\beta  J^2}{4} \biggl(
   q_{\rmcc}^2+q_{\rmss}^2+2 q_{\rmcs}^2
 -\frac{1}{2} \biggr)\nonumber \\
&& -\frac{1}{\beta} \int Dx \int Dy \log \int d \phi M(\phi | x,y),\\
 M(\phi| x,y)&=& \exp\biggl[
   \frac{\beta ^2 J^2}{4}   - \frac{\beta ^2 J^2}{2}\biggl( q_{\rmcc} \cos ^2 \phi
 + q_{\rmss} \sin ^2 \phi \biggr)
- \beta ^2 J^2 q_{\rmcs} \sin \phi  \cos \phi \nonumber \\
&& + \beta J
\sqrt{\frac{q_{\rmcc} q_{\rmss}-q_{\rmcs}^2}{q_{\rmss}}}     
       \cos \phi \ x
       + \beta J \biggl(\frac{ q_{\rmcs}}{\sqrt{q_{\rmss}}}\cos \phi
       +\sqrt{q_{\rmss}} \sin \phi \biggr)y \biggr].
 \eea
 We solved the SPEs for $q_{\rmcc}, q_{\rmss}$ and $q_{\rmcs}$ and found
  the solution with $q_{\rmcc}=q_{\rmss}, \  q_{\rmcs}=0$. Thus, we assume these relations and obtain
\bea
 \bar{f}_{\rm RS} &=& -
   \frac{\beta  J^2}{2}  q_{\rmcc}^2
 -\frac{1}{\beta} \int Dx \int Dy \log \int d \phi M(\phi | x,y),\\
 M(\phi| x,y)&=& \exp\biggl[
  - \frac{\beta ^2 J^2}{2} q_{\rmcc}
 + \beta J \sqrt{q_{\rmcc}} \biggl(       \cos \phi \ x
        + \sin \phi \  y \biggr) \biggr],
 \eea
 where we omit irrelevant constants.
 Now, we introduce the polar coordinates, $x=r \cos \theta, y=r \sin \theta$.
 Then, we have
 \bea
\bar{f}_{\rm RS} &=& -
   \frac{\beta  J^2}{2}  q_{\rmcc}^2
   -\frac{1}{\beta} \frac{1}{2 \pi} \int_0 ^\infty dr e^{-\frac{1}{2}r^2} r \int_0 ^{2 \pi}
 d \theta  \log \int d \phi M(\phi | r, \theta),\\
 M(\phi| r, \theta)&=& e^{
  - \frac{\beta ^2 J^2}{2} q_{\rmcc}
 + \beta J \sqrt{q_{\rmcc}} r \cos (\phi - \theta)}.
 \eea
 By performing integration, we have
 \bea
\bar{f}_{\rm RS} &=& -
   \frac{\beta  J^2}{2}  q_{\rmcc}^2
   -\frac{1}{\beta} \int_0 ^\infty dr e^{-\frac{1}{2}r^2} r
 \log  \biggl(
 2 \pi I_0(\beta J \sqrt{q_{\rmcc}} r) e^{  - \frac{\beta ^2 J^2}{2} q_{\rmcc}} \biggr) \nonumber \\
 &=&
 - \frac{\beta  J^2}{2}  q_{\rmcc}^2
 +  \frac{\beta  J^2}{2}  q_{\rmcc}
   -\frac{1}{\beta} \int_0 ^\infty dr e^{-\frac{1}{2}r^2} r
 \log  \biggl(
 2 \pi I_0(\beta J \sqrt{q_{\rmcc}} r)  \biggr).
  \eea
 In general, $I_n(x)$ is the  modified Bessel function of the $n$th kind.
 The SPE becomes
 \bea
 - \beta  J^2  q_{\rmcc}
 +  \frac{\beta  J^2}{2}
   -\frac{1}{\beta} \int_0 ^\infty dr e^{-\frac{1}{2}r^2} r
   \frac{I_1(\beta J \sqrt{q_{\rmcc}} r)}{I_0(\beta J \sqrt{q_{\rmcc}} r)}
  \beta J r \frac{1}{2\sqrt{q_{\rmcc}}}=0.
 \eea
 Since the spin glass order parameter is
  $q=2q_{\rmcc}$,  we obtain
  \bea
  q &=& 1
  -      \frac{k_{\rm B} T}{ J} \sqrt{\frac{2}{q}}
  \int_0 ^\infty dr r^2  e^{-\frac{1}{2}r^2} 
  \frac{I_1(\frac{J}{k_{\rm B}T} \sqrt{\frac{q}{2}}r)}
{I_0(\frac{J}{k_{\rm B}T} \sqrt{\frac{q}{2}}r)}.
\eea
This is nothing but the equation for $q$ derived by Sherrington and Kirkpatrick\cite{sk}.
\end{document}